# Multi-Beam Energy Moments of Measured Compound Ion Velocity Distributions

**M. V. Goldman,[1] D. L. Newman,[1] J. P. Eastwood,[2] G. Lapenta,[3]**
**J. L. Burch[4] and B. Giles[5]**

[1]Physics Department, University of Colorado at Boulder.
MG Orcid 0000-0001-8039-9143; DN Orcid 0000-0003-0810-1204
[2]The Blackett Laboratory, Imperial College London, London SW7 2AZ, UK.
Orcid 0000-0003-4733-8319
[3]Department of Mathematics, KU Leuven, University of Leuven, Celestijnenlaan 200B, 2001, Leuven, Belgium. Orcid 0002-3123-4024
[4]Southwest Research Institute, San Antonio, TX
[5]NASA Goddard Space Flight Center, Greenbelt, MD

Martin V. Goldman (goldman@colorado.edu)

## Abstract

Compound ion distributions, $f_i(\mathbf{v})$, have been measured with high-time resolution by NASA's Magnetospheric Multi-Scale Mission (MMS) and have been found in reconnection simulations. A compound distribution, $f_i(\mathbf{v})$, consisting, for example, of essentially disjoint pieces will be called a *multi-beam* distribution and modeled as a sum of "beams," $f_i(\mathbf{v}) = f_1(\mathbf{v}) + ... + f_N(\mathbf{v})$. Velocity moments of $f_i(\mathbf{v})$ are taken beam by beam and summed. Such *multi-beam moments* of $f_i(\mathbf{v})$ have advantages over the customary *standard velocity moments* of $f_i(\mathbf{v})$, for which there is only one *mean flow velocity*. For example, the *standard* thermal energy moment of a pair of equal and opposite cold particle beams is non-zero even though each beam has zero thermal energy. We therefore call this thermal energy pseudothermal. By contrast, a *multi-beam* moment of two or more beams has no pseudothermal energy.

We develop three different ways of approximating as a sum of beams and finding multi-beam moments for both a multi-beam $f_i(\mathbf{v})$ measured by MMS in the dayside magnetosphere during reconnection and a multi-beam $f_i(\mathbf{v})$ found in a PIC simulation of magnetotail reconnection. The three methods are:

- A *visual* method in which the velocity centroid of each beam is estimated, and the beam densities are determined self-consistently,
- A k-means method in which particles in a particle-representation of $f_i(\mathbf{v})$ are sorted into a minimum energy configuration of N (= k) clusters,
- A nonlinear *least squares* method based on a fit to a sum of N kappa functions.

Multi-beam energy moments are calculated and compared with standard moments for the thermal energy density, pressure tensor, thermal energy flux (heat plus enthalpy fluxes), bulk kinetic energy density, ram pressure and bulk kinetic energy flux. Applying this new formalism to real data demonstrates in detail how multi-beam techniques provide new insights into the properties of observed space plasmas.

## 1 Introduction

Particle velocity distributions, f(**v**), in space plasmas have been measured with increasing resolution over the past two decades by means of electrostatic analyzer instruments (Fazakerley, et al., 1998). The launch of the Fast Plasma Instrument (FPI) (Pollock et al., 2016) on board the Magnetosphere Multi-Scale (MMS) Mission (Burch, et al, 2016) now enables the full 3D electron and ion distribution to be measured over unprecedented time-intervals of 30 ms and 150 ms cadence respectively (or 7.5 ms and 37.5 ms in certain operating modes (Rager, et al, 2018).

A common and useful method of analyzing particle velocity distributions has been to take *standard* velocity moments of f(**v**). Moments enable the study of f(**v**) at a given spatial point and time (or over a space-time average) in terms of *fluid* variables, such as flow velocity, energy density, temperature, pressure, etc. The meaning of fluid moments is well understood and not controversial for contiguous and effectively single-peaked velocity distributions, f(**v**). However, recently measured electron and ion velocity distributions (Burch, et al, 2016, Eastwood, et al, 2014) can be effectively *disjoint* and are not easily interpreted in terms of *standard* velocity moments. We shall refer to such f(**v**) as *compound* or "*multi-beam"* velocity distributions.

An alternative formal way of taking *multi-beam moments* of *multi-beam* velocity distributions, of form f(**v**) = $f_1$(**v**) + ... + $f_N$(**v**), has been developed recently (Goldman, et al 2020). A multi-beam moment of f(**v**) results from taking a standard moment of each of the "beams," $f_1$(**v**), ..., $f_N$(**v**), and then summing over the beams. Multi-beam moments are often more easily interpreted than standard moments; they don't contain *pseudothermal* energy, such as that present in the standard thermal moment of an f(**v**) consisting of two equal and opposite cold beams. The *pseudothermal part* of any standard thermal moment is defined as the difference between the standard thermal moment and its multi-beam counterpart.

In the present paper we examine how this formalism can be applied to real data, using a multi-beam ion velocity distribution, $f_i$(**v**), measured during dayside magnetopause reconnection (Burch, et al, 2016) and an $f_i$(**v**) found in a Particle-In-Cell (PIC) tail reconnection simulation (Eastwood, et al, 2014; Goldman et al, 2016). For a given f(**v**), we calculate both the standard and multi-beam moments, thereby identifying and eliminating the pseudothermal parts of standard thermal moments. Although each of the three different methods developed in this paper gives (slightly different) multibeam moments, we find they provide reliable means to calculate multi-beam moments of compound distribution functions.

The key step in taking multi-beam moments is to find a way to approximate a measured ion multi-beam $f_i$(**v**) as a sum of N effective "beam" contributions, $f_i$(**v**) = $f_1$(**v**) + ... + $f_N$(**v**). Three different methods are developed for doing this.

- A *visual* method in which the velocity centroid of each beam is estimated
- A *k-means* method in which particles in a particle-representation of $f_i$(**v**) are sorted into a minimum energy configuration of N (= k) clusters
- A nonlinear *least squares* method based on a sum of N kappa functions.

The organization of the remainder of this paper by Section is as follows:

*2. Review of standard and multi-beam moments of particle velocity distributions, f(**v**)*

Kinetic theory definitions of standard velocity moments are reviewed and distinguished from their multi-beam counterparts. *Pseudothermal* moments are defined as differences between standard and multi-beam *thermal moments.* They are expressed entirely in terms of beam densities and velocities. Pseudothermal parts of the standard thermal energy moments are shown to be equivalent to bulk kinetic energy deficiencies of the same moments.

*3. Visual method of finding multi-beam moments of a multi-beam f(**v**)*

Visually estimating the centroid velocities of N = 4 assumed beams enables one to find the beam densities and multibeam moments of a velocity distribution $f(\mathbf{v}) = f_1(\mathbf{v}) + f_2(\mathbf{v}) + f_3(\mathbf{v}) + f_4(\mathbf{v})$ whose standard moments are known. The visual method can also be applied (with constraints) to N = 2, 3 or 5 or more beams.

*4. k-means and least squares methods of approximating f(**v**) and finding multi-beam moments*

Two additional methods are described for decomposing a given velocity-space distribution, f(**v**), into a sum of N beams (without any constraint on N). The first method, typically referred to as "*k*-means," divides the given f(**v**) into $k = N$ disjoint "clusters" in such a way that the thermal energy summed over the *k* clusters is a minimum. The second method uses a nonlinear "least squares" algorithm to minimize the cumulative square of the difference between a chosen distribution, f(**v**), and the corresponding value specified by a parameterized model comprising a superposition of populations.

*5. Multibeam moments of an ion distribution, f(**v**), from a PIC simulation of magnetotail reconnection*

The three methods of finding multibeam moments are applied to an ion velocity distribution, f(**v**), taken from a PIC simulation of magnetotail reconnection. In most cases N is assumed to be 4, and in one case, 2. For the k-means and least-squares methods, the N beam distributions are shown explicitly. The multibeam and standard energy densities are compared, as are the energy density flux vectors. Ion heat fluxes are found to be negligible.

*6. Multibeam moments of ion distribution measured by MMS during dayside reconnection*

Multibeam moments are found by each method for an ion velocity distribution, f(**v**), *measured* by MMS during magnetopause reconnection. Three-dimensional isosurface contour plots are used to illustrate f(**v**) and the beams. Multibeam and standard energy densities are compared, as are the energy density flux vectors. Ion heat fluxes are found to be non-negligible.

*7. Overall Summary and Significance*

A summary is presented of the three methods for approximating an f(**v**) measured in a PIC simulation of magnetotail reconnection and an f(**v**) measured by MMS in the dayside magnetosphere by a sum of N beams. The practical significance of multibeam moment analysis for magnetic reconnection events is explained.

## 2. Review of standard and multi-beam moments of particle velocity distributions, f(v)

In Goldman, et al (2020) it is shown how to define and take multi-beam moments of a compound f(v) from the densities and velocities of N beams in the approximation f(v) = $f_I$ (v) + ... + $f_N$ (v). For completeness, we summarize salient details here.

### 2.1 REVIEW OF STANDARD MOMENTS IN TERMS OF KINETIC THEORY

First, we review definitions and notation used here for *standard* moments of f(**v**):

**Table 1** *Standard* velocity moments of f(v)

| **Name of standard moment** | **Symbol** | **Definition** (dimensional units) |
|---|---|---|
| Number density | n | $\int d^3\mathbf{v} f(\mathbf{v})$ |
| Number density flux | n**u** | $\int d^3\mathbf{v} f(\mathbf{v})\mathbf{v}$, **u** = flow velocity |
| Bulk kinetic energy density | $U_{bulk}$ | $nmu^2/2$ |
| Thermal energy density (***Frame invariant)*** | $U_{therm}$ | $(m/2)\int d^3\mathbf{v} f(\mathbf{v})(\delta v)^2$, $\delta\mathbf{v} = \mathbf{v}-\mathbf{u}$ |
| Total (undecomposed) energy density | U | $U_{bulk} + U_{thermal} = (m/2)\int d^3\mathbf{v} f(\mathbf{v})v^2$ |
| Bulk kinetic energy flux vector | $\mathbf{Q}_{bulk}$ | $n\mathbf{u}mu^2/2$ |
| Heat flux vector (***Frame invariant)*** | $\mathbf{Q}_{heatflux}$ | $(m/2)\int d^3\mathbf{v} f(\mathbf{v})(\delta\mathbf{v})\cdot(\delta v)^2$, $\delta\mathbf{v} = \mathbf{v}-\mathbf{u}$ |
| Enthalpy flux vector | $\mathbf{Q}_{enthalpy}$ | $(m/2)\int d^3\mathbf{v} f(\mathbf{v})[\mathbf{u}(\delta v)^2 + 2\mathbf{u}\cdot(\delta\mathbf{v})(\delta\mathbf{v})]$ |
| Thermal flux vector | $\mathbf{Q}_{therm}$ | $\mathbf{Q}_{heatflux} + \mathbf{Q}_{enthalpy}$ |
| Total (undecomposed) energy flux | **Q** | $\mathbf{Q}_{bulk} + \mathbf{Q}_{thermal} = (m/2)\int d^3\mathbf{v} f(\mathbf{v})\mathbf{v}v^2$ |
| Pressure tensor **(Frame invariant)** | **P** | $m\int d^3\mathbf{v} f(\mathbf{v})(\delta\mathbf{v})(\delta\mathbf{v})$ |
| Ram pressure tensor | $\mathbf{P}_{RAM}$ | $nm\mathbf{uu}$ |
| Stress tensor (undecomposed) moment | **T** | $\mathbf{P}_{RAM} + \mathbf{P} = m\int d^3\mathbf{v} f(\mathbf{v})\mathbf{vv}$ |
| Thermal energy density in terms of pressure | $U_{therm}$ | Tr **P**/2 |
| Enthalpy flux in terms of pressure | $\mathbf{Q}_{enthalpy}$ | $\mathbf{u}$ Tr $\mathbf{P}/2 + \mathbf{u}\cdot\mathbf{P}$ |

It is evident from Table (1) that each standard *undecomposed* moment, **T,** U or **Q** is the sum of the corresponding (coherent) bulk and (incoherent) thermal moments.

## 2.2 MULTI-BEAM MOMENTS

There are at least two ways to take multi-beam moments of a multi-beam f(**v**). One, already mentioned in the *Introduction*, is to take a standard moment of each of the terms (beams) in $f(\mathbf{v}) = f_1(\mathbf{v}) + ... + f_N(\mathbf{v})$ and then sum. The other is to use only the densities, $n_1$, ... $n_N$ and the centroid velocities of the N beams $\mathbf{u}_1$, ... $\mathbf{u}_N$ as described in Goldman, et al (2020). From that paper we reproduce, in Table (2), expressions for the multi-beam thermal moments in terms of the "pseudothermal" parts of the thermal moments, $\Delta U_{therm}$, $\Delta \mathbf{Q}_{therm}$, and $\Delta \mathbf{P}$ using standard moments and only the N partial beam densities, $\eta_j = n_j/n$ and centroid velocities, $\mathbf{u}_j$. Note the dimensionless units introduced in the middle column in Table 2.

As explained in Goldman, et al (2020), the so-called *"pseudothermal"* parts of standard moments are equivalent to differences between *multi-beam bulk* moments and the corresponding *standard bulk* moments. For example, $\Delta U = [U_{therm} - U^{MB}_{therm}]$ is *also* equal to $[U^{MB}_{bulk} - U_{bulk}]$, because of the equivalence of multi-beam and standard undecomposed moments, $U^{MB} = U$. This allows us to express, for example,

(1a) $U^{MB}_{therm} = U_{therm} - \Delta U = U - U^{MB}_{bullk}$

The pseudothermal parts of the moments are therefore determined from the bulk properties (requiring only the $\mathbf{u}_j$'s and $\eta_j$'s) by using Table (2). If the standard moment is known (e.g., by the usual moment-taking procedure) the multibeam moments are determined,

One can think of the *pseudothermal* part of a standard thermal energy moment as a *deficiency* in the bulk kinetic part of the corresponding standard bulk moment. The *pseudo-thermal* energy density and the *bulk kinetic energy deficiency* are equal to each other whether they are scalars (as for energy density moments), tensors (as for pressure moments), or vectors (as for energy density flux moments).

**Table 2: Formulas for determining multi-beam moments** from standard thermal moments, beam densities and beam centroid velocities (dimensionless unit scaling factor in 3rd column**)**

| **f(v) moment** | **Pseudothermal part of standard moment** | **Units** | **Multibeam moment** |
|---|---|---|---|
| Energy Density | $\Delta U \equiv \frac{1}{u_n^2}\left[\sum_{j=1}^{N}\eta_j u_j^2 - u^2\right]$ | $\frac{mnu_n^2}{2}$ | $\boxed{U^{MB}_{therm} = U_{therm} - \Delta U}$ |
| Pressure tensor | $\Delta \mathbf{P} \equiv \frac{1}{u_n^2}\left[\sum_{j=1}^{N}\eta_j \mathbf{u}_j \mathbf{u}_j - \mathbf{u}\mathbf{u}\right]$ | $mnu_n^2$ | $\boxed{\mathbf{P}^{MB} = \mathbf{P} - \Delta \mathbf{P}}$<br>$Tr\mathbf{P} = U_{therm}, \quad Tr\mathbf{P}^{MB} = U^{MB}_{therm}$ |
| Energy Flux vector | $\Delta \mathbf{Q} = \frac{1}{u_n^3}\left[\sum_{j=1}^{N}\eta_j \mathbf{u}_j u_j^2 - \mathbf{u}u^2\right]$ | $\frac{mnu_n^3}{2}$ | $\boxed{\mathbf{Q}^{MB}_{therm} = \mathbf{Q}_{therm} - \Delta \mathbf{Q}}$ $\quad \mathbf{Q}_{therm} \equiv \mathbf{Q}_{enthalpy} + \mathbf{Q}_{htflux}$,<br>$\mathbf{Q}_{enthalpy} \equiv \hat{\mathbf{u}} Tr\mathbf{P} + 2\hat{\mathbf{u}} \cdot \mathbf{P}$ |

## 2.3. DEPENDENCE OF MULTI-BEAM MOMENTS ON NUMBER OF BEAMS, N

As explained in Goldman, et al (2020), the so-called *"pseudothermal"* parts of standard moments are equivalent to differences between *multi-beam bulk* moments and the corresponding *standard bulk* moments. For example, $\Delta U = [U_{therm} - U^{MB}_{therm}]$ is *also* equal to $[U^{MB}_{bulk} - U_{bulk}]$, because of the equivalence of multi-beam and standard undecomposed moments, $U^{MB} = U$. This allows us to express, for example,

(1a) $U^{MB}_{therm} = U_{therm} - \Delta U = U - U^{MB}_{bullk}$

The pseudothermal parts of the moments are therefore determined from the bulk properties (requiring only the $\mathbf{u}_j$'s and $\eta_j$'s) by using Table (2). If the standard moment is known (e.g., by the usual moment-taking procedure) the multibeam moments are determined,

One can think of the *pseudothermal* part of a standard thermal energy moment as a *deficiency* in the bulk kinetic part of the corresponding standard bulk moment. The *pseudo-thermal* energy density and the *bulk kinetic energy deficiency* are equal to each other whether they are scalars (as for energy density moments), tensors (as for pressure moments), or vectors (as for energy density flux moments).

In the construction of multi-beam moments using any of the methods described in this paper, one must first decide the on the *number of beams*, N, in the representation $f(\mathbf{v}) = f_1(\mathbf{v}) + ... + f_N(\mathbf{v})$. The visual method is simplest when $N \leq 4$, whereas there is no restriction on N for the k-means and least squares methods. A recent study based on a sum of Maxwellian $f_j(\mathbf{v})$ suggests that the most statistically likely values of N are low single digits (Dupuis, et al, 2020).

A choice of larger N tends to give a smaller $U^{MB}_{therm}$ and a larger pseudothermal energy, $\Delta U$. This is consistent with the fact that, in the limit of large N, we recover the *kinetic* result that there are only discrete particles: If N is equal to the number of particles there can be no temperature or thermal energy density. These are strictly *fluid* concepts.

In this paper we will mainly use N = 4 (i.e., four beams) for all methods, to provide a clear illustration of their implementation and to enable comparisons. We will also include one N = 2 two-beam example for the visual method. The two-beam case gives the largest value of $U^{MB}_{therm}$ and the smallest pseudothermal energy (ie., the smallest deficit of bulk kinetic energy). The pseudothermal part of the standard pressure moment is a tensor and the pseudothermal part of the standard thermal energy flux is a vector, so one must consider direction as well as magnitude in the pseudothermal parts. In other words, $\mathbf{Q}^{MB}_{bulk}$ can change in magnitude and direction as beams are added to N. Once again,

(1b). $\mathbf{Q}^{MB}_{therm} = \mathbf{Q} - \mathbf{Q}^{MB}_{bulk}$

## 3. Multi-beam moments of a compound f(v) found from beam velocities; the *visual method.*

In this Section it is explained how to find *multi-beam moments* of f(**v**) if beam velocities are estimated, as in the *visual method*, or otherwise known. The beam densities are fully determined from the beam velocities when there are four or fewer beams ($N \le 4$). Once the beam velocities and densities are known and the standard moment and flow velocity of f(**v**) are also known, Table (2) gives the multibeam moments.

### 3.1 THE VISUAL METHOD

The visual method begins with a visual representation of the measured f(**v**) at a given place and time (or over a narrow space-time average near one place and time). This may be in the form of plots of the three mutually orthogonal *reduced* velocity distributions of f(**v**) or a 3-D isosurface of f(**v**).

By *inspection,* one then estimates the *velocity centroids,* $\mathbf{u}_j$ of each of the "beams," $f_1(\mathbf{v})$, ..., $f_N(\mathbf{v})$, as well as the number of beams, N. As will be shown below, if N = 4, the *densities* of each of the beams, $n_1$, $n_2$, ..... $n_N$ are fully determined by Eqs. 2, below. The density and centroid velocity of beam j are *explicitly defined* via standard zero and first-order velocity moments of $f_j(\mathbf{v})$, $n_j = \int d^3\mathbf{v}\ f_j(\mathbf{v})$ and $n_j\mathbf{u}_j = \int d^3\mathbf{v}\ f_j(\mathbf{v})\mathbf{v}$. With this method, however, the full form of the distributions $f_j(v)$ is neither determined nor needed. The velocity moment integrals are never actually performed on the beams since, after estimating the locations of the beam centroid velocities $\mathbf{u}_j$, the values of the beam densities, $n_j$ are determined from zero and first-order moments of f, as described below. From these and Table 2 the pseudothermal and multibeam moments can be found.

### 3.2 ZERO AND FIRST ORDER MOMENTS OF f(v)

The zero-order-moment of $f(\mathbf{v}) = f_1(\mathbf{v}) + f_2(\mathbf{v}) + ... + f_N(\mathbf{v})$, yields conservation of particles (when integrated over volume),

$$(2a)\quad \int \mathbf{d}^3\mathbf{v} f(\mathbf{v})/n = 1 = \eta_1 + \eta_2 + ... + \eta_N, \quad \eta_j \equiv n_j/n$$

Here, $n_j$ is the density of beam j, n is the cumulative density of the ensemble of beams and $\eta_j = n_j/n$ is the fractional density of each beam. The first-order moment of f(**v**) gives the number density flux. When divided by n, this becomes

$$(2b)\quad \int \mathbf{d}^3\mathbf{v}\mathbf{v} f(\mathbf{v})/n = \mathbf{u} = \eta_1\mathbf{u}_1 + \eta_2\mathbf{u}_2 + ... + \eta_N\mathbf{u}_N,$$

where $u_j$ is the centroid velocity of the $j^{th}$ beam and **u** is the (weighted) mean velocity of the ensemble of beams (i.e. the standard flow velocity of f(**v**)). Each beam velocity, $\mathbf{u}_j$, in the weighted mean, **u**, is multiplied by the fractional beam density, $\eta_j$. Eqs. (2) shows that the fractional densities, $\eta_j$ and beam velocities, $\mathbf{u}_j$ are not independent.

### 3.3 FOUR DENSITIES FOUND FROM FOUR BEAM VELOCITIES

Eqs (2) constitute four Eqs. When fitting a measured multi-beam f(**v**) with exactly four beams (N = 4) the number of fractional densities, $\eta_j$, to be determined is also four, so a solution for the $\eta_j$ exists if the determinant of Eqs. (2) does not vanish. Let the beam velocities be **a**, **b**, **c**, and **d**, where the components of each beam are given by,

$$\mathbf{a} \equiv \{u_{1x}, u_{1y}, u_{1z}\}, \ \mathbf{b} \equiv \{u_{2x}, u_{2y}, u_{2z}\}, \ \mathbf{c} \equiv \{u_{3x}, u_{3y}, u_{3z}\}, \ \mathbf{d} \equiv \{u_{4x}, u_{4y}, u_{4z}\}$$

In practical applications, the implicit $\{v_x, v_y, v_z\}$ coordinate system may be simulation coordinates, or, for satellite data, GSE, or field-aligned, $\{v_{||}, v_{\perp 1}, v_{\perp 2}\}$.

Equations (2) may be expressed in terms of 4-dimensional matrix eqns. A matrix, M, acts on the four-vector, $\boldsymbol{\eta} = \{\eta_1, \eta_2, \eta_3, \eta_4\}$ to yield a four-vector composed of the three components of **u** and a fourth component equal to 1. Assuming M has a non-zero determinant, the inverse gives the four-vector, $\boldsymbol{\eta}$:

$$\text{(3a)} \quad \begin{pmatrix} \eta_1 \\ \eta_2 \\ \eta_3 \\ \eta_4 \end{pmatrix} = M^{-1} \begin{pmatrix} u_x \\ u_y \\ u_z \\ 1 \end{pmatrix}, \quad M = \begin{pmatrix} a_x & b_x & c_x & d_x \\ a_y & b_y & c_y & d_y \\ a_z & b_z & c_z & d_z \\ 1 & 1 & 1 & 1 \end{pmatrix}$$

Thus, the four fractional beam densities, $\{\eta_1, \eta_2, \eta_3, \eta_4\}$ can be evaluated by choosing ion beam velocities, **a**, **b**, **c**, **d** and knowing the mean ion flow velocity **u.** The values of the four fractional beam densities, $\{\eta_1, \eta_2, \eta_3, \eta_4\}$ will depend on the values of the four fitted beam velocities and the known value of **u**.

In applying this method, one must be very careful to choose beam velocities which guarantee that all four $\eta_j$ satisfy $1>\eta_j > 0$. If this condition is satisfied **u** will lie within the velocity-tetrahedron whose four vertices are at the four points **a**, **b**, **c**, **d** and if it is not, **u** will lie outside the tetrahedron. The beam velocities must also be chosen so that none of the eigenvalues of the multi-beam pressure tensor are negative.

### 3.4 THREE DENSITIES FOUND FROM THREE BEAMS (N = 3)

In order to approximate f(**v**) by three beams, **a**, **b** and **c** one must reduce the degrees of freedom in Eqs. (2) so that the solution for the etas is not overdetermined. One way to accomplish this is to choose the three beams so that the velocity plane containing the triangle {**a**, **b**, **c**} also contains the standard flow velocity, **u**. If the in-plane coordinates are $v_{x'}$ and $v_{y'}$ and the normal to the plane is in the $v_{z'}$ direction, then **a**, **b**, **c** and **u** will have only x′ and y′ components and Eqs, (2) and its solution for the three beam densities may be written in matrix form as,

$$(4a).\quad \begin{pmatrix} \eta_1 \\ \eta_2 \\ \eta_3 \end{pmatrix} = M^{-1} \cdot \begin{pmatrix} u_{x'} \\ u_{y'} \\ 1 \end{pmatrix}, \quad M \equiv \begin{pmatrix} a_{x'} & b_{x'} & c_{x'} \\ a_{y'} & b_{y'} & c_{y'} \\ 1 & 1 & 1 \end{pmatrix}$$

A more practical approach is to guarantee that one of the beams (say, **c**) is in the plane in $\{v_x, v_y, v_z\}$-space defined by the "known" velocity vectors **a**, **b**, and **u**. The equation of this plane is $\mathbf{A}\cdot\mathbf{v} = |\mathbf{u}|$, where the dimensionless vector, **A**, is given by the solution to the three equations, $\mathbf{A}\cdot\mathbf{a} = |\mathbf{u}|$, $\mathbf{A}\cdot\mathbf{b} = |\mathbf{u}|$ and $\mathbf{A}\cdot\mathbf{u} = |\mathbf{u}|$:

$$(4b)\quad \mathbf{A} = P^{-1} \begin{pmatrix} |\mathbf{u}| \\ |\mathbf{u}| \\ |\mathbf{u}| \end{pmatrix}, \quad P \equiv \begin{pmatrix} a_x & a_y & a_z \\ b_x & b_y & b_z \\ u_x & u_y & u_z \end{pmatrix}$$

In order to be in the same plane, **c** must satisfy $\mathbf{A}\cdot\mathbf{c} = |\mathbf{u}|$. Taking the dot product of **A** with $\mathbf{a}\eta_1 + \mathbf{b}\eta_2 + \mathbf{c}\eta_3 = \mathbf{u}$ shows that the condition $\eta_1 + \eta_2 + \eta_3 = 1$ is automatically satisfied. Hence, we may replace Eqn. (4a), in which the points **a**, **b**, **c**, and **u** are all expressed in the two-dimensional velocity space $\{v_{x'}, v_{y'}\}$, by an equivalent form in three-dimensional velocity space:

$$(4c)\quad \begin{pmatrix} \eta_1 \\ \eta_2 \\ \eta_3 \end{pmatrix} = M_3^{-1} \begin{pmatrix} u_x \\ u_y \\ u_z \end{pmatrix}, \quad M_3 = \begin{pmatrix} a_x & b_x & c_x \\ a_y & b_y & c_y \\ a_z & b_z & c_z \end{pmatrix},$$

$$(4d)\quad \mathbf{A}\cdot\mathbf{a} = \mathbf{A}\cdot\mathbf{b} = \mathbf{A}\cdot\mathbf{c} = \mathbf{A}\cdot\mathbf{u} = |\mathbf{u}|$$

It follows from Eqn. (4d) that **A** is orthogonal to the vectors $(\mathbf{a} - \mathbf{b})/|\mathbf{u}|$ and $(\mathbf{a} - \mathbf{c})/|\mathbf{u}|$. Therefore, **A** is normal to the plane containing the dimensionless $\mathbf{a}/|\mathbf{u}|$-$\mathbf{b}/|\mathbf{u}|$-$\mathbf{c}/|\mathbf{u}|$ triangle.

We now demonstrate that if the three etas are between zero and one, the point **u** will lie inside the triangle formed by **a**, **b** and **c**, and if one of the etas is negative, **u** will be outside the triangle. Since $\eta_1 + \eta_2 + \eta_3 = 1$, no sum of two etas can be negative or the third will be greater than one. It remains to be shown that no single eta can be negative or greater than one.

The weighted mean, **m(a,b)**, of points **a** and **b** may be written as,

$$\mathbf{m}(\mathbf{a},\mathbf{b}) \equiv \frac{\eta_1\mathbf{a} + \eta_2\mathbf{b}}{(\eta_1 + \eta_2)} = \mathbf{a} + g(\eta_1/\eta_2)(\mathbf{b} - \mathbf{a}), \quad \text{where, } g(z) = \frac{1}{(1+z)}$$

For $0 \le \eta_1 \le 1$ and $0 \le \eta_2 \le 1$, $g(\eta_1/\eta_2)$ ranges from a minimum of 0 (at $\eta_2 = 0$, $\eta_1 \neq 0$) to a maximum of 1 (at $\eta_1 = 0$, $\eta_2 \neq 0$), so that **m(a,b)** is on the line interval between the points **a** and **b**. To see this more clearly, we choose a coordinate system such that the origin is at **a** and the line interval between **a** and **b** lies on the x-axis. Then, along this axis, m = g(z)b, where $z = \eta_1/\eta_2$ ranges from 0 to b as z ranges from 0 to infinity. For these z-values, m lies on the line interval ab.

For $-1 < z < 0$, $g(z) > 1$ and for $z < -1$, $g(z) < 0$. In other words, for $-1 < \eta_1/\eta_2 < 0$, m lies on the same line but beyond b and for $\eta_1/\eta_2 < -1$, m is negative, so it lies on the same line but before a. Therefore, if either $\eta_1$ or $\eta_2$ is negative, m lies *outside* the interval ab. To complete the proof, we take the weighted mean of **m**(**a**,**b**) with **c.** This weighted mean is equal to **u**.

$$\mathbf{m}(\mathbf{m}(\mathbf{a},\mathbf{b}),\mathbf{c}) = \frac{(\eta_1+\eta_2)\mathbf{m}(\mathbf{a},\mathbf{b})+\eta_3\mathbf{c}}{(\eta_1+\eta_2)+\eta_3} = \frac{\eta_1\mathbf{a}+\eta_2\mathbf{b}+\eta_3\mathbf{c}}{\eta_1+\eta_2+\eta_3} = \mathbf{u}$$

$$= \mathbf{m}(\mathbf{a},\mathbf{b}) + g(z)[\mathbf{c}-\mathbf{m}(\mathbf{a},\mathbf{b})], \quad z = \frac{(\eta_1+\eta_2)}{\eta_3}$$

The last equality says that **u** lies on the same line as **m(a,b)** and **c**. Since $(\eta_1+\eta_2) > 0$, z is positive only for $\eta_3 \geq 0$, and g is between 0 and 1, so that **u** lies on the line interval between **m**(**a,b**) and **c**. If, in addition, $\eta_1$ and $\eta_2$ are both $\geq 0$, **m**(**a,b**) is on the line interval between **a** and **b** and **u** lies *within* the triangle abc. If any of the three etas is negative, either **m**(**a,b**) is outside of the line interval between **a** and **b** or **u** is outside the interval between **m**(**a,b**) and **c**. Either way, **u** is outside the triangle abc. This completes the demonstration.

In addition to the condition on the etas it again must be the case that the chosen velocities **a**, **b**, and **c** are such that the eigenvalues of the pressure matrix must be positive.

After **a**, **b**, and **c** are chosen subject to these constraints, the pseudo-potential and multibeam moments can be found from **a**, **b**, **c**, $\eta_1$, $\eta_2$ and $\eta_3$.

### 3.5 TWO DENSITIES FOUND FROM TWO BEAM VELOCITIES (N = 2)

For two-beam approximations to f(**v**) the reduction of Eqns. (2) to two degrees of freedom requires that the beam velocities **a** and **b** lie on a line in velocity space which also contains the flow velocity **u**. Along this line Eqns (2) and its solution become,

(4e) $$\begin{pmatrix} a & b \\ 1 & 1 \end{pmatrix}\begin{pmatrix} \eta_1 \\ \eta_2 \end{pmatrix} = \begin{pmatrix} u \\ 1 \end{pmatrix}, \quad \begin{pmatrix} \eta_1 \\ \eta_2 \end{pmatrix} = \frac{1}{a-b}\begin{pmatrix} u-b \\ a-u \end{pmatrix},$$

where a, b and u are the locations along the line of the tips of the vectors **a**, **b** and **u**.

The estimates of beam velocities **a** and **b** are constrained to be such that $\eta_1 > 0$ and $\eta_2 > 0$. It is easy to see from the last equality in Eqn. (5) that this requires ordering along the line of either a > u > b or b > u > a. Hence, u must be sandwiched between a and b if both etas are to be positive.

### 3.6 FOUR DENSITIES FROM FIVE BEAMS (N = 5)

For five or more beams Eqns. (2) is *under*determined for the five etas. However, if the fractional density, $\eta_5$, and velocity, $\mathbf{u}_5$ of the fifth beam are determined by other means, then the same matrix, M, used above in Eqn. (3) may be employed, together with the source term now containing $\eta_5$ and $\mathbf{u}_5$ to find the other four fractional densities:

$$(5a)\qquad \begin{bmatrix}\eta_1\\ \eta_2\\ \eta_3\\ \eta_4\end{bmatrix} = M^{-1}\cdot\begin{bmatrix}u_x-\eta_5 u_{5x}\\ u_y-\eta_5 u_{5y}\\ u_z-\eta_5 u_{5z}\\ 1-\eta_5\end{bmatrix}$$

To bring this into the standard form of Eqn. (3a), divide by (1- $\eta_5$) to obtain

$$(5b)\qquad \begin{pmatrix}\mu_1\\ \mu_2\\ \mu_3\\ \mu_4\end{pmatrix} = M^{-1}\begin{pmatrix}w_x\\ w_y\\ w_z\\ 1\end{pmatrix},\quad \mu_j=\frac{\eta_j}{(1-\eta_5)},\quad \mathbf{w}=\frac{\mathbf{u}-\eta_5\mathbf{u}_5}{(1-\eta_5)}$$

One use might be for when a feature such as a high energy "halo" is present in f(**v**) and treated as a fifth, split-off beam or *"pedestal."*

### 3.7 THREE DENSITIES FROM FOUR BEAMS

The same logic as above may be applied to N = 4 beams when the velocity and fractional density of the fourth beam, say, **d** and $\eta_4$, are already fully determined (e.g., by taking standard moments of beam four). The task, then, is to estimate the velocities of beams **a**, **b** and **c** and determine their fractional densities $\eta_1$, $\eta_2$ and $\eta_3$. As before, an example is when a feature such as a high energy "halo" is present in f(**v**) and treated as a fourth, split-off beam or *"pedestal"* (Sec. 4.4). It is then evident that **a**, **b**, **c** and (**u -** $\eta_4$**d**) must all lie in the same plane and Eqns. (2) and their solutions for $\eta_1$, $\eta_2$ and $\eta_3$ may be written as,

$$(5c)\qquad \begin{pmatrix}\mu_1\\ \mu_2\\ \mu_3\end{pmatrix} = M_3^{-1}\begin{pmatrix}w_x\\ w_y\\ w_z\end{pmatrix},\quad \mu_j=\frac{\eta_j}{(1-\mu_4)},\quad \mathbf{w}=\frac{\mathbf{u}-\mathbf{d}\eta_4}{(1-\mu_4)}$$

$$(5d)\qquad \mathbf{A}\cdot\mathbf{a}=\mathbf{A}\cdot\mathbf{b}=\mathbf{A}\cdot\mathbf{c}=\mathbf{A}\cdot\mathbf{w}=|\,\mathbf{u}\,|.$$

Provided the standard moments and standard flow velocity of f(**v**) are known, the multibeam moments are then found from the four beam velocities and four densities using Table (2).

## 4. K-means and least square methods of approximating f(v) and finding multibeam moments

In addition to the *visual method*, we have found two complementary methods for decomposing a given velocity-space distribution, f(**v**), into a specified number of populations (e.g., beams).

- The first method, typically referred to as "***k*-means**," divides the given f(**v**) into $N = k$ disjoint "clusters" in such a way that the thermal energy summed over the $N$ clusters is a minimum. The k-means method is a widely used technique, well-known in image analysis.
- The second method uses a nonlinear "**least squares**" algorithm to minimize the cumulative square of the difference between f(**v**) (or a function of f(**v**)) and the corresponding value specified by a parameterized model comprising a superposition of populations.

In each case, we employ algorithms from the MATLAB [R] analysis package; with comparable algorithms implemented in other software (e.g., Mathematica [R]). The methodology is described more fully in the subsections below.

The two "given" velocity distributions, f(**v**), to be used later in this paper will be

- from a PIC simulation (Eastwood, et al, 2014) (Sect. 5)
- from the MMS mission (Sect. 6)

One important way in which the *k*-means and least-squares algorithms differ is in how the data is represented. While the least-squares algorithm operates directly on f(**v**) in 3D velocity space, the *k*-means algorithm requires a collection of velocity-space coordinates of discrete equally weighted *particles* distributed according to f(**v**). Our discussion of the two decomposition models will assume the satellite and simulation data are in the necessary form, with a description of how the data is brought into the required form deferred until the end of this section.

### 4.1 K-MEANS DECOMPOSITION

The starting point for the k-means decomposition is a set of $N_p$ *particles* of uniform weight, each with velocity $\mathbf{v}_i$, for $i = 1, ..., N_p$, distributed according to the velocity distribution function $f(\mathbf{v})$ as described below. The object is to divide the $N_p$ particles into N = *k clusters*. In the *continuum* picture this is equivalent to $f(\mathbf{v}) = f_1(\mathbf{v}) + ... + f_k(\mathbf{v})$, where none of the $f_j(\mathbf{v})$'s are overlapping.

To be consistent with the visual method of beam determination discussed above, we will apply the method with $N = 4$, although it works for N equal to any integer. The k-means algorithm finds the 'center' velocity, $\mathbf{u}_j$, of cluster j, with $j = 1, .., k_N$, labelling each of the $N$ clusters. It also yields the number, $N_j$, of *particles* associated with each cluster. The set of $N_j$ particles in cluster j is associated with one (and only one) center velocity, $\mathbf{u}_j$. The $N_p$ particles are divided into $k$ disjoint sets (clusters) of $N_j$ particles each, with,

$$(6a)\quad \sum_{j=1}^{N} N_j = N_p$$

In addition, the k-means algorithm determines the $k$ center velocities, $\mathbf{u}_j$, so as to satisfy the following conditions:

$$(6b)\ \mathbf{u}_j = \frac{1}{N_j}\sum_{i=1}^{N_j}\mathbf{v}_{ji}, \quad \text{and}$$

$$(6c)\ \sum_{j=1}^{k}\sum_{i=1}^{N_j}\left(\mathbf{v}_{ji}-\mathbf{u}_j\right)^2 = \text{ local minimum}, \quad \text{where, } \mathbf{v}_{ji} = \text{i}^{\text{th}} \text{ partice in j}^{\text{th}} \text{ cluster}$$

Eqn. (6b) states that each cluster center, $\mathbf{u}_j$, is the *centroid* of particle velocities in that cluster.

The condition in Eqn. (6c) is equivalent to the requirement that the total thermal energy of particles relative to their respective cluster centers be a local minimum. The left side of Eqn. (6c) is essentially the *multi-beam* thermal energy, based on the standard $U_{therm}$ in Table (3).

$$U_{therm}^{MB} = \frac{m}{2}\sum_{j=1}^{k}\int d^3\mathbf{v} f_j(\mathbf{v})\left(\mathbf{v}-\mathbf{u}_j\right)^2$$

From the derived expression, $U_{therm}^{MB} = U_{therm} - \Delta U$ it is clear that the k-means procedure will simultaneously *maximize* the *pseudothermal* energy, ΔU, as defined in Table 2.

The k-means algorithm is nonlinear and performs an iterative search for the *k* velocities $\mathbf{u}_j$, yielding a local minimum of the total multibeam thermal energy while meeting the requirement that the cluster centers also be the cluster centroids. There is therefore no guarantee that the solution is a *global* minimum. By repeating the search with different initial guesses, one can ascertain whether the local minimum is likely close to a global one. For the examples presented in this study, this multibeam thermal energy appear to be close to the global minimum with high likelihood.

The fractional density in each cluster is simply the fraction $N_j/N$ of all particles in cluster *j*. Furthermore, because each center is the centroid of the respective cluster it immediately follows that the standard bulk flow velocity, **u**, satisfies particle density flux conservation,

$$(6d)\ \mathbf{u} = \sum_{j=1}^{k}\eta_j\mathbf{u}_j$$

Because the *k*-means algorithm identifies the cluster center for each of the original *N* particles the full pressure tensor $\mathbf{P}_j$ and heat-flux, $\mathbf{Q}_{j\text{-heatflux}}$, can be computed for each cluster, j. Therefore, terms such as $\mathbf{Q}_{j-heatflux}$ can be determined for each cluster as well as for the sum over all clusters $\mathbf{Q}^{MB}{}_{heatflux}$. This is an advantage over the visual method which does not yield the pressure tensors and heat fluxes of individual beams. Another advantage to this method is that it should be easy to set up automatically.

### 4.2 NONLINEAR LEAST-SQUARES DECOMPOSITION

The nonlinear least-squares (NLSQ) decomposition attempts to find an optimal representation of given velocity distribution function f(**v**) as a superposition of simple analytic distributions functions, $g_j(\mathbf{v})$. Here, we will use "kappa" distributions for our base set, $g_j(\mathbf{v})$, j = 1, ... , N.

The density, bulk velocity, temperature tensor, and index κ of each $g_j(\mathbf{v})$ are taken as parameters to be determined through the fitting algorithm. The "data" to be fit is, in general, a scalar function of the 3D f(**v**) on a cubic Cartesian grid of dimension $N_{vx} \times N_{vy} \times N_{vz}$ along the three orthogonal basis directions.

The MATLAB algorithm we use assumes a 1D array of data values. This entails simply concatenating the 3D data into a one-dimensional data vector. Since the computation time and memory demand increase with the number of data points (as well as the number of model parameters), we subsample our distribution function before performing the fit. A grid of $N_g = 93^3 \approx 5.7 \times 10^5$ data points is found to afford a reasonable trade off between velocity-space resolution and computational demand.

The model distribution we will use to approximate the satellite or simulation data in this study will be a superposition of triaxial kappa distribution, for which the tri-Maxwellian distribution is a limiting form as $\kappa \to \infty$. The essential feature of such distributions (in their center-of-mass frame) is that each $g_j(\mathbf{v})$ is constant on nested ellipsoidal surfaces of uniform shape and orientation. Specifically, each $g_j(\mathbf{v})$ depends on velocity only through the dimensionless scalar function, h(**v**), defined as follows:

$$(7)\quad \mathrm{h}(\mathbf{v}) \equiv \frac{nM}{2}\left(\mathbf{v}^T \mathsf{P}^{-1}\mathbf{v}\right) = \frac{M}{2}\left(\frac{v_1^2}{T_1} + \frac{v_2^2}{T_2} + \frac{v_3^2}{T_3}\right),$$

where $(v_1, v_2, v_3)$ are the components of velocity, **v**, in the coordinate system for which the pressure tensor is diagonal (it can always be diagonalized, due to the symmetry property $P_{ij} \equiv P_{ji}$). The mass of the species being considered is denoted by M. The scalars, $T_i$, for $i = 1, 2, 3$ are the eigenvalues of $\mathsf{P}/n$ (i.e., the diagonal elements of $\mathsf{P}/n$ in that coordinate system). Note that $\mathbf{v}^T$ and **v** have the components of the velocity in the original coordinate system and are expressed as row and column vectors, respectively.

Each kappa distribution is a function of velocity, **v**, characterized by 11 parameters: the density, $n$ (scalar); the local mean velocity vector, **u**, (three components); the symmetric temperature tensor, $\mathsf{P}$, (which has six independent components); and the scalar index $\kappa$. The explicit form of each parameterized kappa distribution, $f_{\mathrm{kappa}}$ takes the following form:

$$(8)\quad f_{kappa}(\boldsymbol{v}; n, \mathbf{u}, \mathsf{P}, \kappa) \equiv \frac{n\Gamma(\kappa+1)/\Gamma(\kappa-1/2)}{\pi(2\kappa-3)^{3/2}\left[T_1T_2T_3\right]^{1/2}/M^{3/2}}\left(1+\frac{h(\boldsymbol{v}-\mathbf{u})}{\kappa-3/2}\right)^{-(\kappa+1)}$$

,

Note $f_{kappa}$ has the usual physical units, $n/v^3$.

The *fitting* distribution g(**v**) = $\sum_{j=1}^{N} g_j(\mathbf{v})$ consists of a superposition of $N$ independent kappa functions, each labeled by subscript j. Since $g_j(\mathbf{v}) = f_{kappa}(\mathbf{v}; n_j, \mathbf{u}_j, \mathsf{P}, \kappa_j)$, the fitting distribution, g(**v**) has $11N$ free parameters.

The MATLAB NLSQ algorithm allows an upper bound, lower bound, or both to be set on any parameter. We constrain $n$ to be positive for each component and restrict $\kappa$ to the range $2 \le \kappa \le \kappa_{\max}$, but do not impose any constraints on the other parameters.

The upper bound ($\kappa_{\max} = 50$ for the PIC simulation and $\kappa_{\max} = 10$ for the FPI data) differs relatively little from a Maxwellian (especially for $\kappa_{\max} = 50$) except well out on the power-law ($g \sim v^{-2\kappa_{\max}}$) tail. The lower bound $\kappa = 2$ is larger than the minimum value $\kappa = 3/2$ where the normalization of the second velocity moment (i.e., **P**) of the distribution breaks down. Nevertheless, with $g(v) \sim v^{-4}$ for larger $v$ higher moments of the distribution such as the heat flux, may diverge. This, however, does not affect the validity of fitting with a small index such as $\kappa = 2$, because the fit is restricted to the domain of the finite velocity-space data grid. Outside of this domain, the tails of any physical distribution will likely decrease more rapidly with $v$. Note, however that the heat flux can remain finite even with a $v^{-4}$ tail on the velocity distribution, provided the tails become sufficiently *symmetric* at high velocity (see Section on tri-Maxwellians in Goldman, et al, 2020))

In its simplest form, the NLSQ algorithm varies the model parameters to minimize the cumulative square error $\sum_{n=1}^{N_g} |f(\mathbf{v}_n) - g(\mathbf{v}_n)|^2$, where the integer subscript n labels grid points ranging from 1 to $N_g$.

Note that any scalar function of the distribution can be used instead. This is particularly useful for modeling the observed FPI distributions where the cumulative square error due to differences in the tail of the distribution are insignificant and therefore cannot effectively influence the value of $\kappa$ in the model. However, by first taking the logarithm of the distribution [both the observed $f(\mathbf{v})$ and the model $g(\mathbf{v})$], the controlling feature of the distribution shifts from the local maxima to the power-law tails. A hybrid approach in which both $f(\mathbf{v})$ and $\log_{10}[f(\mathbf{v})]$ are used in the NLSQ fitting process is described in connection with the modeling of the actual FPI data.

### *4.3 PREPARATION OF SATELLITE AND SIMULATION DATA*

As described in the preceding sections, the *k*-means algorithm requires the particle distribution be represented by an ensemble of equally weighted discrete particles while the NLSQ algorithm requires the distribution function be specified on a 3D Cartesian grid. While the output of a PIC simulation might be directly usable as input to the *k*-means algorithm, the iPic3D code used in the reconnection simulation described here uses particles of unequal weight and therefore the simulation output requires additional processing. Since we need the velocity distribution on a Cartesian velocity-space grid for the NLSQ method, we first generate such a distribution from either the PIC simulation output or from the FPI distribution skymap and then generate an ensemble of equally weighted particles from the resulting distribution grid. A brief description of the methodology follows:

#### 4.3.1 FROM IPIC3D PARTICLE DATA TO A CARTESIAN f(**v**)

The ensemble of *weighted* particles culled from the iPic3D reconnection simulation comprise the totality of particles located in a square $0.5d_{i0}$ on a side centered at the location where the two trajectory bundles in Fig. 3e and 3f of Goldman, et al, (2020) cross ($x = 120d_{i0}$ and $y = 14d_{i0}$). There are $\sim 3 \times 10^4$ macro-particles in the sample region.

The velocity distribution function is created by adding the weight of each particle to the cumulative value for the cubic cell containing the particle's velocity in a pre-defined Cartesian velocity-space grid. Since only $\sim 1.6 \times 10^4$ particles have a significant weight, the resulting distribution has relatively large statistical fluctuations. We apply five iterations of nearest-neighbor smoothing to reduce the fluctuation level as a visualization aid. The reduced distributions in the figures are based on the smoothed distributions.

#### 4.3.2 FROM FPI SKYMAPS TO A CARTESIAN f(**v**)

The FPI skymaps take the form of a distribution function over a 16×32×32 grid in $\theta$, $\phi$, E space where E is the logarithmically spaced energy in eV. While the value of the distribution function can be interpolated directly onto a Cartesian velocity-space grid, we employ an intermediate step whereby the logarithm of the distribution on a shell of constant E is fit to a superposition of spherical harmonics $Y_{lm}(\theta,\phi)$ up to a predetermined order $l_{\max}$. In the present study we use $l_{\max} = 13$ for which there are $(l_{\max} + 1)^2 = 196$ parameters (i.e., $Y_{lm}$ amplitude coefficients) to fit the 32 × 16 = 512 angular data point (subject to a floor value in those angular bins where there are zero counts). The reconstruction of the distribution on each energy shell from the corresponding $Y_{lm}(\theta, \phi)$ is then used for the interpolation onto the predetermined Cartesian velocity-space grid.

#### 4.3.3 FROM CARTESIAN f(**v**) TO AN ENSEMBLE OF EQUALLY WEIGHTED PARTICLES

The procedure for generating the desired ensemble of $N_p$ equally weighted particles distributed according to the 3D velocity distribution function on a Cartesian grid consists of the following steps:

1. Map the 3D Cartesian velocity-space grid of dimension ($N_{vx}$, $N_{vy}$, $N_{vz}$) into a 1D vector of length $N_g = N_{vx} \times N_{vy} \times N_{vz}$. The gridded velocity distribution $f(v_x,v_y,v_z)$ then maps onto the data vector $f_j$, $j$=1...,$N_g$.
2. Construct the *cumulative* distribution vector, $F_j = \sum_{i=1}^{j} f_i$, with $F_{\text{tot}} = F_{Ng}$ being the sum over all grid points of $f$.
3. Generate a set of $N_p$ random numbers uniformly distributed between 0 and $F_{\text{tot}}$. These values will be used to identify the velocities of the $N_p$ particles in the target ensemble.
4. Interpolate each random number onto the vector $F_j$ defined in step 2, rounding up to the next higher integer $j$.
5. Invert the mapping defined in step 1 from $j$ back to the gridded 3D velocity ($v_x$, $v_y$, $v_z$).
6. Add a random velocity perturbation to each particle so that its final velocity is randomly distributed with the corresponding cubic grid cell.

Note that we are using $v_x$, $v_y$, and $v_z$ as generic velocity components. For the 2D PIC magnetotail reconnection simulation studied in Sec. (6), the initial state is a Harris current sheet of finite thickness in y, with *current* in the (out-of-plane) negative-z-direction, producing reversed magnetic field lines in the ±x directions. For this PIC simulation, the velocity components, { $v_x$, $v_y$, $v_z$ ,} are in these simulation coordinates. (This differs from the GSE convention, in which x is the same but y and -z are exchanged)

For the case of the dayside MMS FPI-measured velocity distribution, $v_x$, $v_y$, and $v_z$ will be interpreted as magnetic field-aligned velocities $\{v_{\|}, v_{\perp 1}, v_{\perp 2}\}$.

It is essential that the cubic cells in the Cartesian velocity-space grid all have the same volume for either case to work without modification.

## 4.4. HIGH ENERGY CLOUDS AND PEDESTALS

Particle distributions in many space plasma environments often contain a suprathermal *halo* or *cloud* population, which can play a significant role when evaluating the multibeam moments of the distribution. Therefore, it is advantageous to have a means to first approximate the total distribution as the sum of two populations: the first representing the suprathermal halo and the second representing the remaining *core* population. The multi-component character of the core can then be better approximated after separating out the halo population. This technique will be used in combination with the *k*-means decomposition methodology of Sec. 4.1 and with the visual method. An alternative approach to separating the distribution into a suprathermal halo and (multi-component) core using nonlinear least square decomposition (Sec. 4.2) is discussed in Sec. 6.4. In the context of the *k*-means method, including the halo populations when performing the decomposition can have a significant influence on the velocities of the cluster centers.

This effect can be understood by considering the innermost sum in Eq. (6c): $\sum_{i=1}^{N_j}(v_{ji}-u_j)^2$. Because halo particles can have very large velocities relative to the cluster centers, the square of their distance from these centers $v_{ji}-u_j$ can have a disproportionate influence on the determination of the optimal $u_j$ by the *k*-means algorithm due to the quadratic dependence on the velocity separation. This influence is mitigated by first separating the halo from the core distribution and limiting the *k*-means cluster decomposition (or the visual determination of beam velocities) to the core alone.

The method we employ to separate the halo from the core population is referred to as a *pedestal* method. We consider two variants of this method, although the pedestal analogy applies only to the first of these variants. The procedures are illustrated schematically in Fig. 1. Both variants can be easily implemented as part of an automated analysis of FPI data.

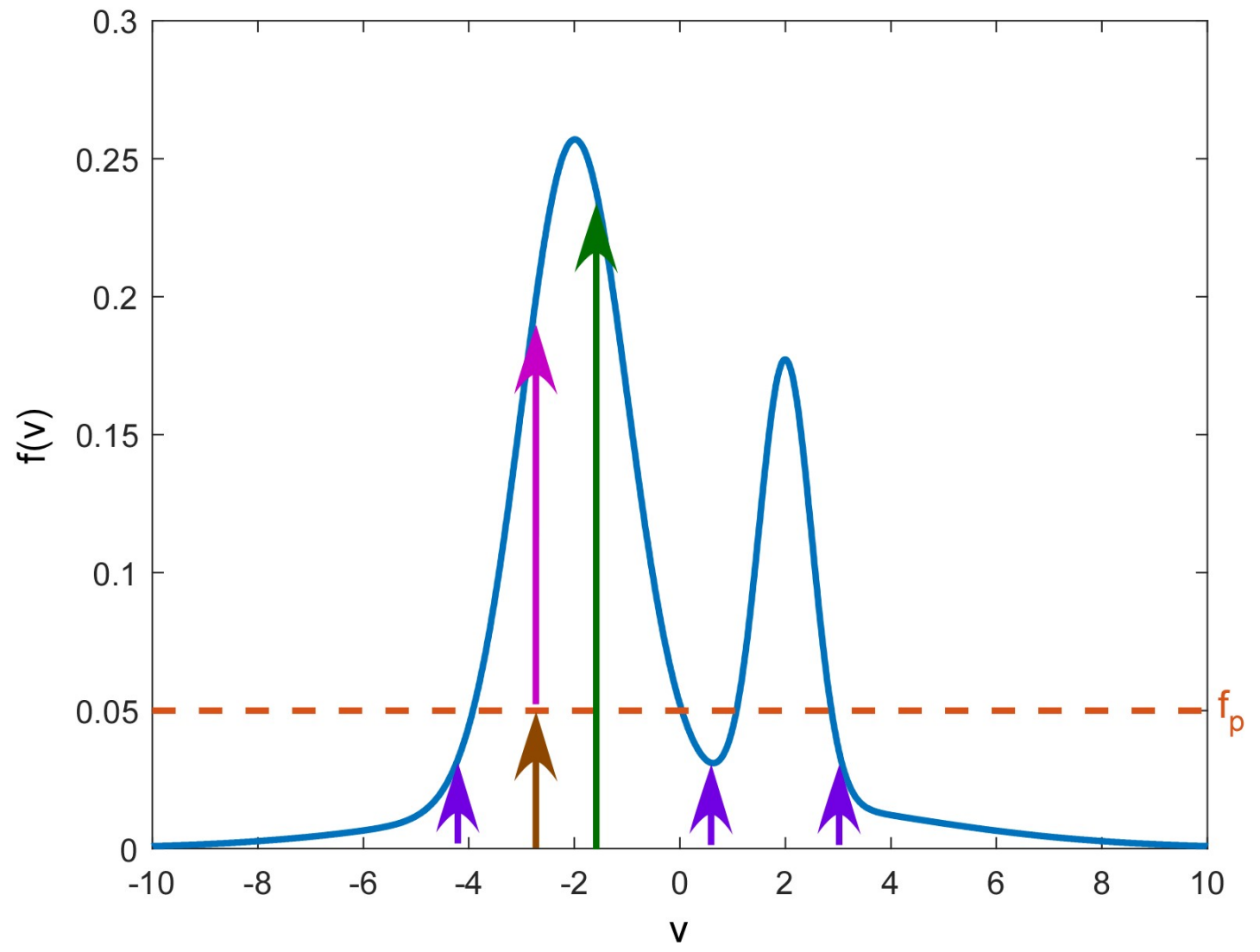

**Figure 1:** Schematic illustrating the separation of a 1-D velocity distribution $f(v)$ [blue curve] into halo and core populations for pedestal value $f_p = 0.05$ [dashed horizonta line]. In both methods, when $f(v) \le f_p$, $f_h(v) = f(v)$ [purple arrows] and $f_c(v) = 0$. When $f(v) > f_p$, the first method sets $f_h(v) = f_p$ [brown arrow] and $f_c(v) = f(v) - f_p$ [magenta arro The second method sets $f_h(v) = 0$ and $f_c(v) = f(v)$ [green ar

The object is to express a velocity distribution $f(\mathbf{v})$ as a sum of halo and core distributions $f(\mathbf{v}) = f_h(\mathbf{v}) + f_c(\mathbf{v})$. By choosing a *pedestal* value of the distribution $f_p$ (independent of $\mathbf{v}$) less than the global maximum of $f(\mathbf{v})$, velocity space can be separated into velocities $\mathbf{v}_\le$ where $f(\mathbf{v}) \le f_p$ and velocities $\mathbf{v}_>$ where $f(\mathbf{v}) > f_p$. The regions $\mathbf{v}_\le$ and $\mathbf{v}_>$ each can consist of either a single connected set of velocities or multiple disjoint sets, with the union of $\mathbf{v}_\le$ and $\mathbf{v}_>$ comprising the entirety of velocity space.

The first method of separating $f(\mathbf{v})$ in to $f_h(\mathbf{v})+f_c(\mathbf{v})$ defines $f_h(\mathbf{v}) = f(\mathbf{v})$ for $\mathbf{v} \in \mathbf{v}_\leq$ and $f_h = f_p$ for $\mathbf{v} \in \mathbf{v}_>$. The second method differs only in that $f_h = 0$ for $\mathbf{v} \in \mathbf{v}_>$. In both cases, $f_c$ is defined as $f_c(\mathbf{v}) = f(\mathbf{v}) - f_h(\mathbf{v})$. These differences are illustrated schematically in Fig. 1 for a 1-D velocity distribution consisting of two beams together with a broader background distribution. For the value of $f_p$ in the figure, $\mathbf{v}_\leq$ consists of three disjoint regions (roughly $v < -4$, $0 < v < 1$, and $v > 3$), while $\mathbf{v}_>$ consists of two disjoint regions (roughly $-4 < v < 0$ and $0 < v < 1$). For the first separation method, $f_h$ follows the minimum of the solid blue and dashed orange curves, which has the appearance of a pedestal upon which $f_c$ sits (motivating the terminology). For the second method, there are gaps in the 1-D $f_h(v)$ where $f(v) > f_p$. These gaps become *holes* or *voids* in $f_h(\mathbf{v})$ in higher dimensions.

For the first separation method (but not the second), $f_h$ and $f_c$ are *not* disjoint when $f(\mathbf{v}) > f_p$ as depicted by the brown and magenta arrows in Fig. 1, which are co-located in velocity. Since the *k*-means algorithm requires an ensemble of equally weighted particles to operate on, the procedure outlined in Sec. 4.3.3 must be modified slightly. Specifically, any particle associated with velocity-space region $\mathbf{v}_>$ is randomly assigned to either $f_h$ or $f_c$ with respective probabilities $f_p/f(\mathbf{v})$ and $[f(\mathbf{v}) - f_p]/f(\mathbf{v})$.

In the application to the FPI ion distribution in Sec. 6.3, the core distribution $f_c(\mathbf{v})$ is subject to a *k*-means decomposition with $k = 3$. These three populations together with the halo distribution $f_h(\mathbf{v})$ are treated as four individual *beams* for the purpose of computing the multibeam moments.

## 5. Multibeam moments of f(**v**) from *PIC simulation* of magnetotail reconnection

A *visual* representation of the ion velocity distribution whose moments are to be taken is the starting point for the visual method. Although not intrinsic to the methods, a visual representation is also helpful in visualizing the results of the k-means and the least squares methods of finding multibeam moments.

In both this Section, which addresses velocity distributions from PIC simulation, and in the next (dealing with MMS-FPI ion velocity measurements during dayside reconnection) two visual representations are employed

- a set of three orthogonal *reduced* 2D contour plots of the ion velocity distribution.
- *3D fly-around surface-contour plot* of the ion velocity distributions

The underlying ion velocity distribution is of necessity somewhat different for each of the three methods of finding multibeam moments. In *the present section* the visual method estimates the effective *beam velocities* by inspection of the ion velocity distributions obtained by processing and smoothing the PIC results (see Section (4.3.1)) to obtain the simulation "box" ion distribution, $f_{box}(\mathbf{v})$. The k-means method uses $f_{box}(\mathbf{v})$ to create the distribution of ion *particles* necessary to employ the k-means method (Sec. (4)). Finally, the least squares method is based on the *least-squares* velocity distribution, $f_{fit}(\mathbf{v})$ *fitted* to $f_{box}(\mathbf{v})$.

The multibeam moments found in this section by each of the three methods assumes that $f_{box}$ or $f_{fit}$ can be interpreted as a sum of beams, $f(\mathbf{v}) = f_1(\mathbf{v}) + ... + f_N(\mathbf{v})$, with N = 4 or 2. The *pseudothermal* parts of the *standard* moments of f(**v**) are found by comparing the standard moments with the multibeam moments in each case. The multibeam moments $U^{MB}_{therm}$ and $\mathbf{Q}^{\mathbf{MB}}_{therm}$ will differ from the standard moments $U_{therm}$ and $\mathbf{Q}_{therm}$ by amounts equal to the pseudothermal part of each standard moment (Table 2). The *standard* moments will differ for each of the three methods for finding multibeam moments since, in general, $f(\mathbf{v}) \neq f_{box}(\mathbf{v}) \neq f_{fit}(\mathbf{v})$. For the PIC example treated in this Section there is little difference between the three different underlying velocity distributions, $f(\mathbf{v})$, $f_{box}(\mathbf{v})$ and $f_{fit}(\mathbf{v})$.

For the FPI case to be addressed in Section (6), the *visual* method is based on the FPI *standard* moments file, with *beams* estimated visually by inspecting the velocity distribution, $f_{box}(\mathbf{v})$, processed from the FPI skymap. The Cartesian velocity coordinate $f_{box}(\mathbf{v})$ will be based on the FPI-measured spherical velocity-coordinate skymap, as described in Section (4). Owing to a high-velocity *"cloud"* or "*halo"* present in the FPI distributions (but not in the PIC distributions) the three standard moments differ more than in the PIC case.

### 5.1 REDUCED ION VELOCITY DISTRIBUTION, f(**v**), FROM **PIC** SIMULATION

The reduced ion velocity distributions of $f_{box}(\mathbf{v})$ for the magnetotail reconnection PIC simulation are shown in Fig. (5), in terms of simulation coordinates, $\{v_x, v_y, v_z\}$, and *reduced* velocity distibutions, $F_a$, $F_b$, $F_c$.

The 2D spatial simulation coordinates are {x, y, z}, where z is out-of-plane. The x-velocity, $\mathbf{v}_x$, is the component of **v** colinear with the initial in-plane magnetic field; the y-velocity, $\mathbf{v}_y$, is in the vertical in-plane direction, and $\mathbf{v}_z$ is out of plane. Field aligned coordinates are not necessary here since the ions are not strongly magnetized in this region.

The *reduced* velocity distibutions, $F_a$, $F_b$, $F_c$ are defined in terms of $\{v_x, v_y, v_z\}$ as,

$$\text{(9a)}\quad F_a(v_x, v_y) \equiv \int_{-\infty}^{\infty} dv_z\, f(v_x, v_y, v_z)$$

$$\text{(9b)}\quad F_b(v_x, v_z) \equiv \int_{-\infty}^{\infty} dv_y\, f(v_x, v_y, v_z)$$

$$\text{(9c)}\quad F_c(v_y, v_z) \equiv \int_{-\infty}^{\infty} dv_x\, f(v_x, v_y, v_z)$$

The three reduced distributions are depicted in three orthogonal planes in Fig. (2), below.

The reduced distributions may not be best-suited for visualizing the dominant beam-like portions of f(**v**). This is because there may be multiple beams along one direction in a reduced velocity distribution compressed by the velocity integration along that direction. A more comprehensive "fly-around" 3D representation of f(**v**) will be used shortly, resulting in different beam centroid velocities and multi-beam moments found from the *visual* method.

Two principal beams are especially apparent in both $F_a(v_x, v_y)$ and $F_b(v_y, v_z)$ in Figs. (2a) and (2c). The third reduced distribution, $F_c(v_x, v_z)$ in Fig. (2b), peaks in a crescent shape, which is harder to interpret (hence, the advantage of a "fly-around" 3D representation of f(**v**)).

Around the pair of peaks is a high-velocity *cloud* or "*halo*." *Clouds* are present and much more extended to high velocities in the MMS f(**v**) to be considered in Section (6). Clouds can have consequences for the heat flux moment which will be discussed later.

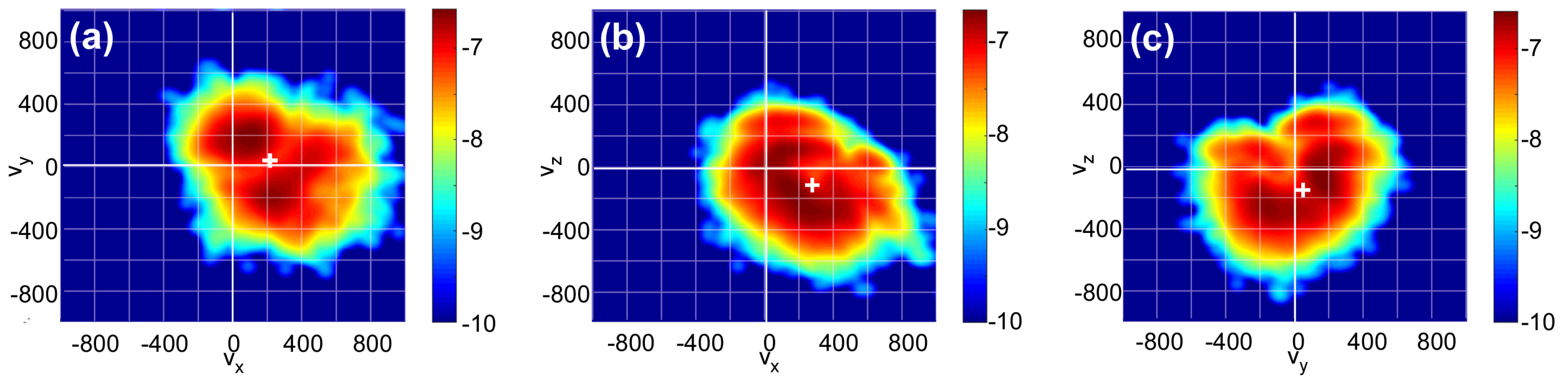

**Figure 2:** Contour plots of logarithm of reduced velocity distributions of the PIC ion f(**v**): (a) $F_a(v_x, v_y)$, (b) $F_b(v_x, v_z)$, and (c) $F_c(v_y, v_z)$ defined in Eqns. (9). Units of velocities and of velocity distributions are arbitrary. The intersection of the horizontal and vertical white lines is the velocity-space origin, {0,0}. Components of the mean flow, **u**, are indicated by the + sign in each of the three planes. Note that the flow is mainly in the $v_x$-directions

### 5.2 ***STANDARD MOMENTS*** *OF f(v) FROM PIC SIMULATION*

In *standard* energy-transport analyses one generally uses the *standard* decomposed moments of f(**v**). The *visual method* determines the *multibeam moments* by subtracting the *pseudothermal moments* from the corresponding *standard moments*, using the equations in Table (2). The standard moments of $f_{box}(\mathbf{v})$ which can be used to find multibeam moments for the PIC simulation are given below in Table (3)

**Table 3:** ***Standard*** **moments of $f_{box}(\mathbf{v})$ from PIC magnetotail reconnection simulation** (Fig. (2)). The standard mean flow velocity, **u** is in arbitrary units. The standard pressure tensor, **P**, standard energy density, U, and standard energy flux vectors, $\mathbf{Q}_{bulk}$, $\mathbf{Q}_{enthalpy}$ and **Q** are in the dimensionless units defined in Table (2), with $\mathbf{u}_n = \mathbf{u}$. Hence, the standard energy flux vectors are all in units of $\mathbf{Q}_{bulk}$ and $|\mathbf{Q}_{bulk}| = 1$. The standard moments **u**, **P,** U and **Q** are all calculated from the same velocity distribution, $f_{box}(\mathbf{v})$.

| **Standard** moments of $f_{box}(\mathbf{v})$ from **TAIL PIC SIMULATION** | x | y | z | Norm, Trace or Eigenvalues |
|---|---|---|---|---|
| Mean flow, **u** | 226 | 10 | -123 | $\lvert\mathbf{u}\rvert$ =257.5 |
| Presure, $P_{\mathbf{x,i}}$ | 0.705 | -0.208 | -0.22 | Tr**P** = $U_{therm}$ = 2.06 |
| Pressure, $P_{\mathbf{y,i}}$ | -0.208 | 0.695 | 0.179 | Eigenvalues(**P**) = |
| Pressure, $P_{\mathbf{z,i}}$ | -0.22 | 0.179 | 0.644 | {1.09, 0.50, 0.45} |
| $\mathbf{Q}_{bulk}$ | 0.88 | 0.039 | -0.48 | $\lvert\mathbf{Q}_{bulk}\rvert$ =1.00 |
| $\mathbf{Q}_{enthalpy}$ | 3.23 | -0.40 | -1.96 | $\lvert\mathbf{Q}_{enthalpy}\rvert$ = 3.80 |
| $\mathbf{Q}_{heatflux}$ | 0.15 | -0.01 | 0.12 | $\lvert\mathbf{Q}_{heatflux}\rvert$ =0.19 |
| $\mathbf{Q}_{therm} = \mathbf{Q}_{enthalpy} + \mathbf{Q}_{heatflux}$ | 3.38 | -0.41 | -1.85 | $\lvert\mathbf{Q}_{therm}\rvert$ = 3.87 |
| $\mathbf{Q} = \mathbf{Q}_{bulk} + \mathbf{Q}_{therm}$ | 4.26 | -0.37 | -2.33 | $\lvert\mathbf{Q}\rvert$ = 4.87 |

Table (3) gives the standard mean flow velocity, **u**, in arbitrary units as well as the components of the pressure tensor and energy flux vectors (bulk, enthalpy and heat fluxes) in appropriate dimensionless units (Table (2), with $\mathbf{u}_n = \mathbf{u}$).

Note that the heat flux is negligible in magnitude compared to the enthalpy. This is because of the absence of high velocity "haloes" in the PIC f(**v**), as seen in Fig. (2). In a later section, MMS-measured velocity distributions in the magnetopause with *extended* velocity haloes will be found, together with large heat fluxes. This association arises because heat fluxes are third order veocity moments which are strongly influenced by higher velocities (i.e., on the *tail* of f(**v**)).

The standard pressure tensor, **P**, is symmetric, so it possesses only six independent matrix elements instead of nine. The trace of the pressure tensor, Tr**P**, is about 2 in these units, and is identically equivalent to the dimensionless standard *thermal energy density*, $U_{therm}$. (Table (1)). Hence there is no need to calculate $U_{therm}$ separately. Moreover, since the *bulk* energy density is 1 in these units, the total standard ion energy density moment is $U = 1 + U_{therm} = 3.06$

The eigenvalues and eigenvectors of the standard presure, **P** (both pseudo and true thermal parts), specify the *thermal asymmetry* of f(**v**). The three eigenvalues of **P** give three generally different pressures, $P_1$, $P_2$ and $P_3$ associated with each of the three *principal axes* defined by the directions of the three orthogonal eigenvectors.

The nature of the *pressure anisotropy* is illustrated in Fig. (3). The red, green and blue vectors point in the directions, $\{\hat{\mathbf{x}}', \hat{\mathbf{y}}', \hat{\mathbf{z}}'\}$, of the three pressure eigenvectors, but have magnitudes equal to the square roots of the corresponding eigenvalues, $\sqrt{P_1}$, $\sqrt{P_2}$, and $\sqrt{P_3}$. These define the semi-major and semi-minor axes of a *pressure ellipsoid* in velocity space. With velocity **v** in units of the magnitude, u, of the flow velocity, **u**,

$$\frac{v_{x'}^2}{P_1} + \frac{v_{y'}^2}{P_2} + \frac{v_{z'}^2}{P_3} = 1$$

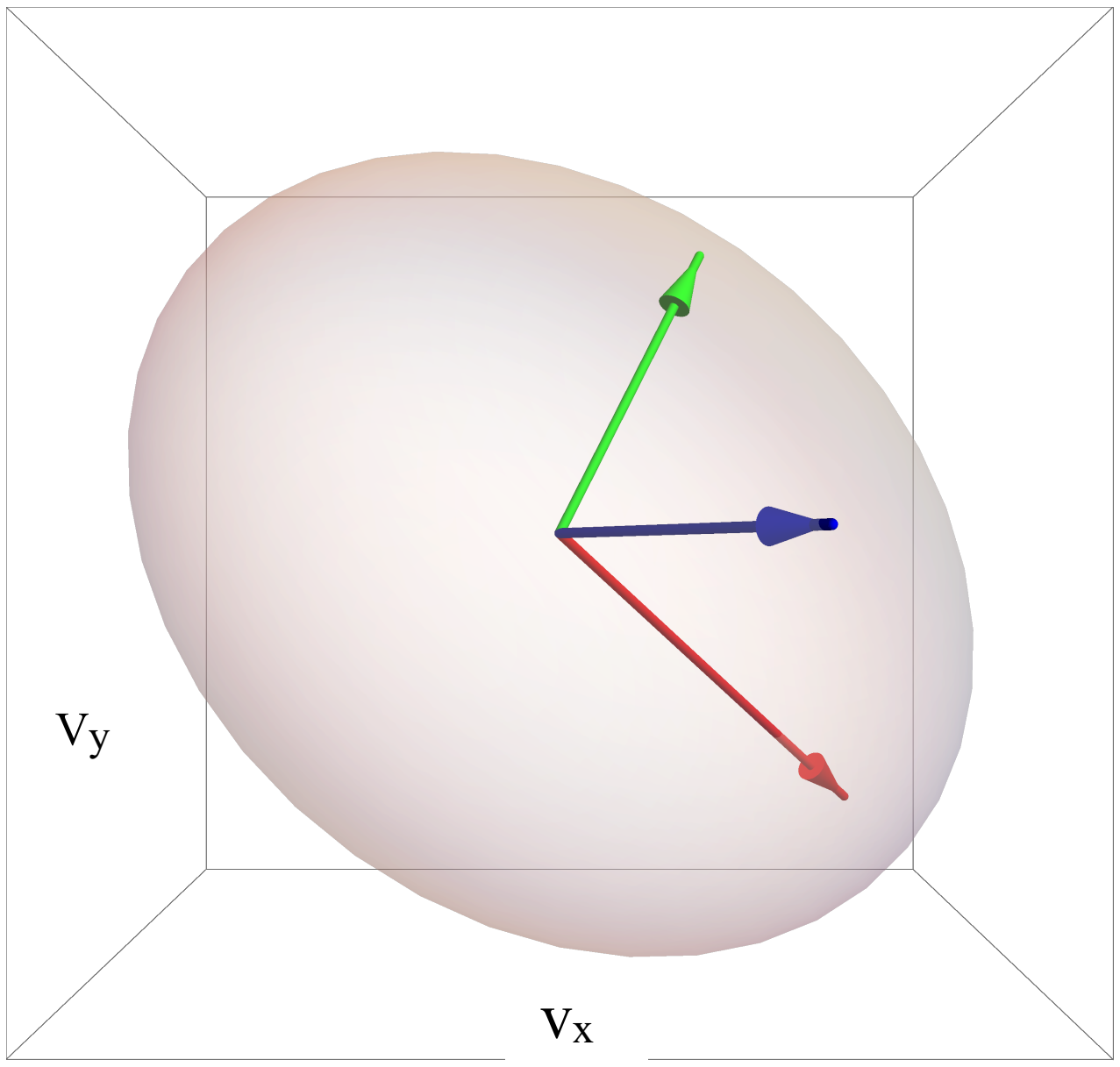

**Figure 3:** Orthogonal eigenvectors of the *standard pressure tensor* for the tail PIC simulation are in the *directions* of the red, blue and green vectors. The colored vectors each have a *magnitude* equal to the square root of the corresponding eigenvalue (Table (3)). Velocities are dimensionless, scaled by the flow speed, u. R, G, and B vectors define semi-major and semi-minor axes of this *pressure ellipsoid.*

The orientation of the ellipsoid in simulation coordinates is oblique to the x-axis. From Table (3), $\sqrt{P_1} = 1.04$, $\sqrt{P_2} = 0.71$ and $\sqrt{P_3} = 0.67$ are the magnitudes of the R, G and B vectors. The magnitudes of the G and B velocity vectors are almost equal, while the magnitude of the R vector is larger, so the pressure ellipsoid in this case is essentially a *prolate spheroid*. Note there is a related *temperature ellipsoid* which can be expressed in terms of the three orthogonal *temperatures,* $T_1 = P_1/n$, $T_2 = P_2/n$, and $T_3 = P_3/n$. This is simply a rescaled version of the pressure ellipsoid.

## 5.3 PIC f(v) MOMENTS FROM 3-D VISUAL METHOD AND K-MEANS METHOD

Fig. (4), below, shows a *3-D velocity contour map visualization* of the PIC ion velocity distribution, f(**v**), whose *reduced* distributions were displayed in Fig. (2). This and other 3-D velocity contour maps will be used to visualize the results of the different methods for approximating f(**v**) as a sum of four ion beams. The beam velocities needed in the visual method are often easier to estimate from a 3-D visualization than from reduced distributions. The four beam velocities found by using the other methods (k-means and least squares) are shown n Figs. (4,5) to compare with the visual method. Multi-beam moments found by each method will be organized in Tables for comparison with each other and with the PIC standard moments given in Table (3). Multi-beam and standard *pressure ellipsoids* associated with each of the three methods are displayed in Figs. (4,5) to give a clearer picture of the *pseudothermal* content of the standard pressure tensors.

The green and yellow surfaces in Fig. (4a) show f(**v**) for the PIC simulation box. The bold magenta plus-mark indicates the location of the standard moment flow velocity, **u**. The other four plus-marks give the estimated velocities of the beams in a four-beam approximation to f(**v**) (see Table (4)). Fig. (4b) shows that the multi-beam pressure ellipsoid (black) found by the visual method is smaller and almost orthogonal to the standard pressure ellipsoid (black).

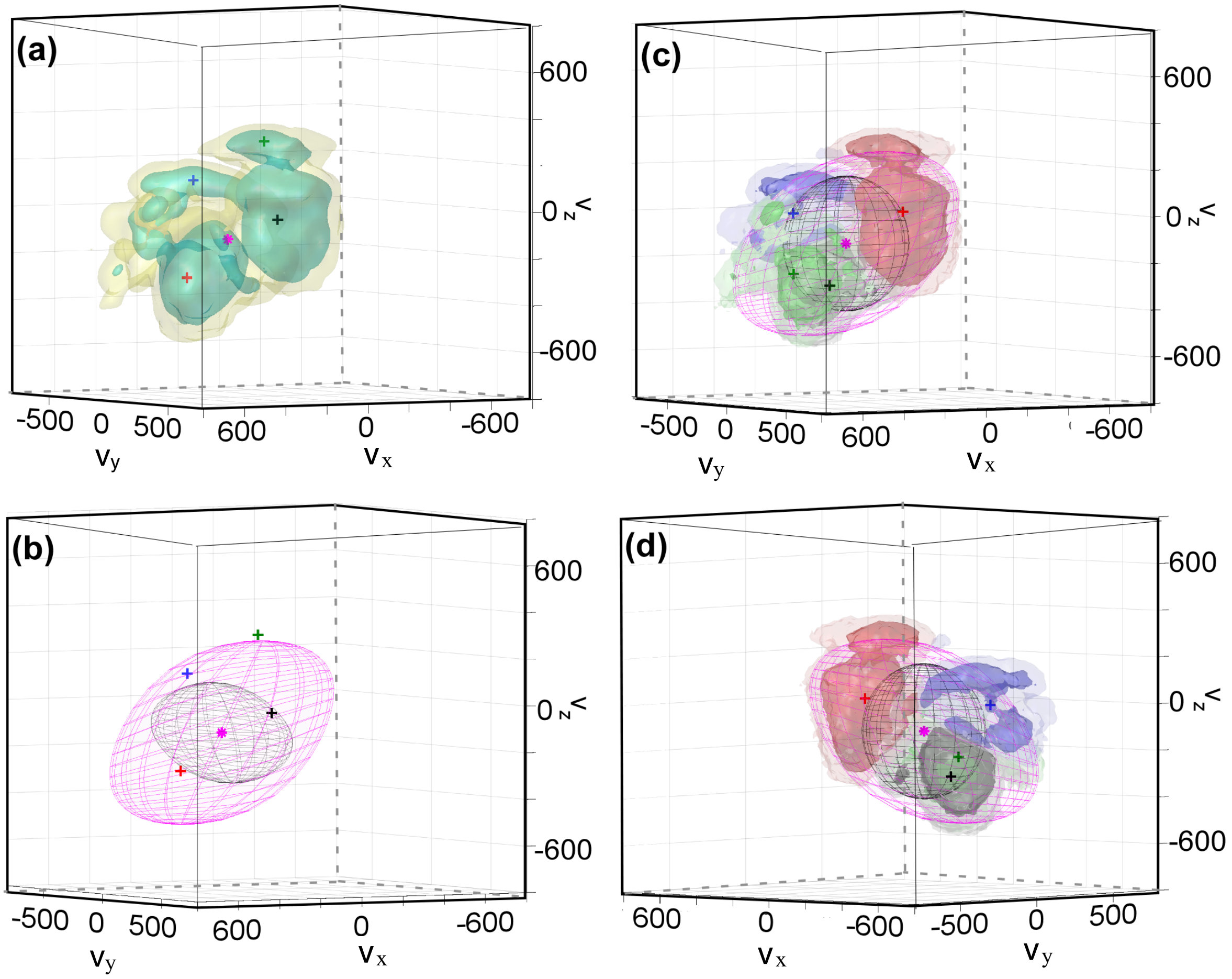

**Figure 4: 3D visualizations for PIC tail simulation:**
**(a)** Surface contour map of PIC velocity distribution, f(**v**). Green surfaces are at 25% of global maximum of f(**v**) and yellow surfaces are at 10%. Magenta star marks the standard flow velocity **u**. Colored plus-marks indicate the four visually estimated beam velocities. (**b**) Pressure anisotropy ellipsoids based on eigenvalues and eigenvectors of pressure tensors for f(**v**). Standard pressure ellipsoid is pink; multibeam pressure ellipsoid is black. Standard and individual beam velocities are again shown. False pressure is contained in space between ellipsoids. (**c**), (**d**) Two surface-contour views (almost 180° apart) showing partition of f(**v**) into four *clusters* obtained by k-means method with four beams (colored black, red, green, and blue). All properties of the four beams are completely determined by k-means.

Figs. (4c) and (4d) are visualizations relevant to the k-means method, which here divides f(**v**) into four non-overlapping "clusters." The clusters are colored green, blue, red, and black in the front and back views of Figs. (4c) and (4d). Also shown in these two figures are the standard pressure ellipsoid (in pink), and the smaller four-beam pressure ellipsoid (in black), oriented almost perpendicular to the multi-beam pressure ellipsoid determined by the four-beam visual method. The four plus marks show the locations of the centroids of the four clusters, which are completely determined by the k-means method rather than the visual method.

Table (4) quantifies the four visually estimated beam velocities in Fig. (3a) and gives the $\eta_j$'s found for each beam from zero and first-order moments, as explained in Section 3.

**Table 4:** Four beam visually-estimated centroid velocities and derived fractional beam densities (etas) from 3D $f_{box}(\mathbf{v})$ for PIC simulation. See Fig. (3a) .

| PIC simulation 3D $f_{box}(v)$ fitted visually with four beams | Beam 1 | Beam 2 | Beam 3 | Beam 4 |
|---|---|---|---|---|
| **x-comp of centroid velocity** | 90 | 350 | 125 | 250 |
| **y-comp of centroid velocity** | 195 | -120 | 140 | -240 |
| **z-comp of centroid velocity** | -35 | -290 | 300 | 120 |
| **Inferred** Density Fraction, $\eta_j$ | 0.399 | 0.466 | 0.054 | 0.081 |

In Table 5 the four-beam moments found by the *visual method* are given in dimensionless units (Table 2). The thermal energy density is ~1, compared to a standard thermal energy density of ~ 2 (Table 3). Thus, about half of the standard thermal energy density is pseudo-thermal; it also, appears as an increase in the four-beam bulk kinetic energy density (not shown). The magnitude of the standard thermal energy flux vector is reduced by about a factor of two. The eigenvalues and eigenvectors of the pressure tensor in the Table were used to construct the four-beam pressure ellipsoid compared with the standard pressure ellipsoid in Fig. (3b).

**Table 5**: Multi-beam pressure tensor, eigenvalues and trace and energy density flux vectors from the visual method applied to 3D $f_{box}(v)$ from PIC simulation. See Fig. (3) above.

| ***PIC Visual Method*** **4-beam moments of** $f_{box}$ | **x** comp. | **y** comp | **z** comp | ***Related scalar/vector*** |
|---|---|---|---|---|
| $P^{MB}_{x,j}$ | 0.477 | 0.075 | 0.026 | $TrP^{MB} = U^{MB}_{therm} = 0.960$ |
| $P^{MB}_{y,j}$ | 0.075 | 0.280 | -0.042 | Eigenvalues of $P^{MB}$ = {0.502, 0.284, 0.174} |
| $P^{MB}_{z,j}$ | 0.026 | -0.042 | 0.202 | |
| $\mathbf{Q^{MB}_{bulk}}$ | 2.420 | -0.606 | -1.593 | $\lvert Q^{MB}_{bulk}\rvert = 2.960$ |
| $\mathbf{Q^{MB}_{therm}}$ | 1.840 | 0.233 | -0.730 | $\lvert Q^{MB}_{therm}\rvert = 1.993$ |
| $\mathbf{Q = Q^{MB}_{bulk} + Q^{MB}_{therm}}$ | 4.260 | -0.373 | -2.323 | $\lvert Q\rvert = 4.866$ |

Shown below in Table (6) are the four-beam velocity moments of f(**v**), as found by the k-means method. Comparing with the standard moments in Table (3), once again the four-beam thermal energy density is smaller than the standard thermal energy density. Here they are in the ratio 2.06 to 0.803 so this method yields even more pseudothermal energy than the visual method.

**Table 6: Multibeam moments found by k-means partitioning of $f_{box}$(v) from PIC simulation:** four-beam pressure tensor moment, $\mathbf{P}^{MB}$, and four-beam energy fluxes $\mathbf{Q}^{MB}_{bulk}$, $\mathbf{Q}^{MB}_{enthalpy}$, $\mathbf{Q}^{MB}_{heatflux}$, $\mathbf{Q}^{MB}_{thermal}$, and $\mathbf{Q}^{MB}$.

| *PIC k-means 4-beam moments of tail simultion f(v)* | x comp. | y comp. | z comp. | ***Related scalar/vector*** |
|---|---|---|---|---|
| $P^{MB}_{x,i}$ | 0.260 | -0.010 | 0.001 | $Tr\mathbf{P}^{MB} = U^{MB}_{therm} = 0.803$ |
| $P^{MB}_{y,i}$ | -0.010 | 0.226 | 0.007 | *Eigenvalues of* $\mathbf{P}^{MB}$ = |
| $P^{MB}_{z,i}$ | 0.001 | 0.00 | 0.318 | {0.318, 0.269, 0.216} |
| $\mathbf{Q}^{MB}_{bulk}$ | 3.079 | -0.375 | -1.725 | $\lvert\mathbf{Q}^{MB}_{bulk}\rvert = 3.55$ |
| $\mathbf{Q}^{MB}_{therm} = \mathbf{Q}^{MB}_{enth} + \mathbf{Q}^{MB}_{htflux}$ | 1.253 | -0.004 | -0.630 | $\lvert\mathbf{Q}^{MB}_{therm}\rvert = 1.40$ |
| $\mathbf{Q}^{MB}_{htflux}$ | -0.013 | -0.002 | -0.006 | $\lvert\mathbf{Q}^{MB}_{htflux}\rvert = 0.014$ |
| $\mathbf{Q}^{MB} = \mathbf{Q}^{MB}_{bulk} + \mathbf{Q}^{MB}_{therm}$ | 4.332 | -0.380 | -2.359 | $\lvert\mathbf{Q}^{MB}\rvert = 4.947$ |

## 5.4 PIC f(**v**) MOMENTS FOUND FROM LEAST SQUARES METHOD

Finally, we address the *least-squares method* of approximating the PIC ion distribution f(**v**) as a sum of four kappa-function ion beams, in accordance with Section 4.2, above.

Figure (5a) shows the results of the four kappa-function beam least-squares fit to the PIC f(**v**) in Fig. (4a). The standard flow velocity, **u**, is marked by the magenta star and the other colored plus-signs mark the velocity centroids of the four beams whose free parameters have been determined from the least-squares fit. Once again, the pink ellipsoid is based on the standard pressure tensor, and the black ellipsoid is based on the four-beam pressure tensor, with pseudothermal pressure in between the two ellipsoids.

Figure (5b) exhibits in a different color each of the four kappa-function beams which are summed in Fig. (5a). The velocity centroid of each beam is indicated by a plus-sign of the same color as the beam. All free parameters have been determined by the least-squares method.

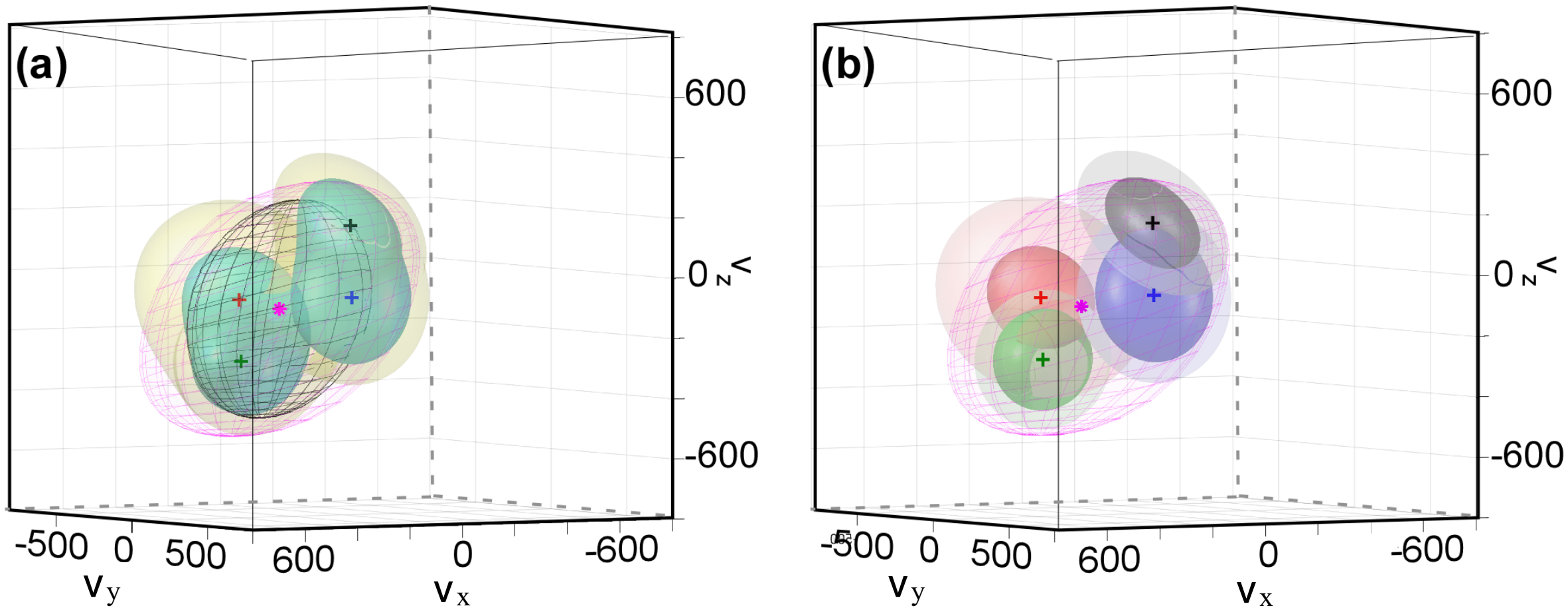

**Figure** 5: 3D visualization of least squares fit to f(**v**) from PIC tail simulation: (a) Surface contour map of least squares four-beam fit, $f_{ls}$(**v**) to PIC velocity distribution, $f_{box}$ (**v**) shown in Fig. (3a). Green surfaces are at 25% of global maximum of $f_{ls}$(**v**) and yellow surfaces are at 10%. Magenta star marks standard flow velocity **u**. Colored plus-marks indicate least squares beam velocity centers. Also shown in (a) are pressure anisotropy ellipsoids based on eigenvalues and eigenvectors of pressure tensors, **P**, for $f_{ls}$(**v**). Standard pressure ellipsoid is pink; multibeam pressure ellipsoid is black. (b) the four different kappa-functions (each in a different color) from the least squares fit shown in (a).

Table (7) gives the four-beam moments of the fitted distribution in Fig. (4a). The four-beam energy density is 1.16, which is slightly more than half of the standard energy density of $f_{box}$ in Table (3). Note that the multi-beam heat flux is zero, while the standard heat flux is non-vanishing. This is because the kappa-function-beams used in the fit have no heat flux, due to symmetry. This does *not* signify that the standard heat flux is entirely pseudothermal.

**Table 7:** Four-beam moments from *least squares* fit, $\mathbf{f_{fit}(v)}$, to $\mathbf{f(v)}$ from PIC simulation: pressure tensor, $\mathbf{P}^{MB}$, multibeam bulk K.E. flux vector, $\mathbf{Q}^{MB}{}_{bulk}$ the thermal energy flux vector, $\mathbf{Q}^{MB}{}_{therm}$ and $Tr\mathbf{P}^{MB} = U^{MB}{}_{thermal}$, in dimensionless units.

| **PIC least squares 4-beam moments** of distribution, f(**v**), fitted to PIC f(**v**) | **x comp.** | **y comp.** | **z comp.** | ***Related scalar/vector*** |
|---|---|---|---|---|
| $P^{MB}{}_{x,i}$ | 0.465 | 0.083 | 0.009 | $Tr\mathbf{P}^{MB} = U^{MB}{}_{therm} = 1.16$ |
| $P^{MB}{}_{y,i}$ | 0.083 | 0.334 | -0.066 | Eigenvalues($\mathbf{P}^{MB}$) = |
| $P^{MB}{}_{z,i}$ | 0.009 | -0.066 | 0.359 | {0.509, 0.395, 0.255} |
| $\mathbf{Q}^{MB}{}_{bulk}$ | 2.107 | -0.290 | -1.290 | $\lvert\mathbf{Q}^{MB}{}_{bulk}\rvert = 2.487$ |
| $\mathbf{Q}^{MB}{}_{htflux}$ | 0 | 0 | 0 | $\mathbf{Q}^{MB}{}_{htflux} = 0$ |
| $\mathbf{Q}^{MB}{}_{therm} = \mathbf{Q}^{MB}{}_{enth} + \mathbf{Q}^{MB}{}_{htflux}$ | 2.085 | 0.045 | -0.904 | $\lvert\mathbf{Q}^{MB}{}_{therm}\rvert = 2.273$ |
| $\mathbf{Q}^{MB} = \mathbf{Q}^{MB}{}_{bulk} + \mathbf{Q}^{MB}{}_{therm}$ | 4.192 | -0.245 | -2.193 | $\lvert\mathbf{Q}^{MB}\rvert = 4.737$ |

A measure of the utility of the fit for finding moments of $f_{box}(\mathbf{v})$ is to compare the *standard* moments of f(**v**) with the *standard* moments of the *fitted* distribution in Fig. (4a). The standard moments of the *fitted* distribution are given in Table (8), below. Consulting Table (3), we see that the standard moments of $f_{fit}(\mathbf{v})$ are acceptably close to the standard moments of f(**v**).

**Table 8:** Standard moments of least squares *fitted* distribution**,** $f_{fit}(\mathbf{v})$ to PIC f(**v**). Note that the standard heat flux of $f_{fit}(\mathbf{v})$ is not zero.

| **PIC standard moments of least sqs. fitted distribution**, $f_{fit}(\mathbf{v})$, for | x | y | z | Norm, Trace or Eigenvalues |
|---|---|---|---|---|
| Presure $P_{\mathbf{x,i}}$ | 0.691 | -0.162 | -0.219 | $Tr\mathbf{P} = U_{therm} = 2.186$ |
| Pressure $P_{\mathbf{y,i}}$ | -0.162 | 0.773 | 0.154 | Eigenvalues(**P**) = |
| Pressure $P_{\mathbf{z,i}}$ | -0.219 | 0.154 | 0.722 | {1.085, 0.615, 0.486} |
| $\mathbf{Q}_{bulk}$ | 0.776 | 0.050 | -0.400 | $\lvert\mathbf{Q}_{bulk}\rvert = 0.875$ |
| $\mathbf{Q}_{heatflux}$ | 0.210 | -0.090 | 0.153 | $\lvert\mathbf{Q}_{heatflux}\rvert = 0.275$ |
| $\mathbf{Q}_{therm} = \mathbf{Q}_{enthalpy} + \mathbf{Q}_{heatflux}$ | 3.415 | 0.295 | --1.793 | $\lvert\mathbf{Q}_{therm}\rvert = 3.868$ |
| $\mathbf{Q} = \mathbf{Q}_{bulk} + \mathbf{Q}_{therm}$ | 4.191 | -0.245 | -2.191 | $\lvert\mathbf{Q}\rvert = 4.737$ |

## 5.5 COMPARISON OF MULTI-BEAM MOMENTS OF PIC f(**v**) FOUND BY DIFFERENT METHODS

A visual comparison of moments helps to understand the similarities and differences among the numerical values of moments of the PIC f(**v**). The pressure ellipsoids in Figs. 4 and 5 already constitute a visual comparison of the pressure tensors. The multi-beam pressure ellipsoids found by the three different methods are generally smaller than the standard pressure ellipsoid of Fig. (3) and embedded within it.

In this Section we display the scalar values of the bulk and thermal energy densities in terms of a bar graph (Fig. 6) and the bulk and thermal energy density *fluxes* in terms of vector diagrams (Fig. 7).

The bars in purple at the bottom of the bar graph show the *standard* energy densities of the PIC f(**v**) (Table 3). The right purple bar is $U_{thermal}$ and the left purple bar is $U_{bulk}$. The sum of two is the undecomposed energy density U, which is also the sum of the multi-beam bulk and thermal energy densities shown in the five bars above the purple standard energy densities. The light orange bars are from the visual four-beam moments (Table 5) and the pink bars are from a different visual four-beam. The yellow bars were found using the visual method with two-beams. The grey and light blue bars at the top are from the k-means method (Table 6) and the least squares method (Table 7). In all cases the pseudothermal energy density is equal to the difference between the standard thermal energy density bar and the multibeam energy density bar. The pseudothermal energy density for all cases is roughly 50% except for the two-beam case, where it is closer to 40%. In all cases the *lost* pseudothermal energy shows up as a *gain* in bulk energy density. The sum of the bulk and thermal energy densities (standard or multibeam) is 3.1 in all cases.

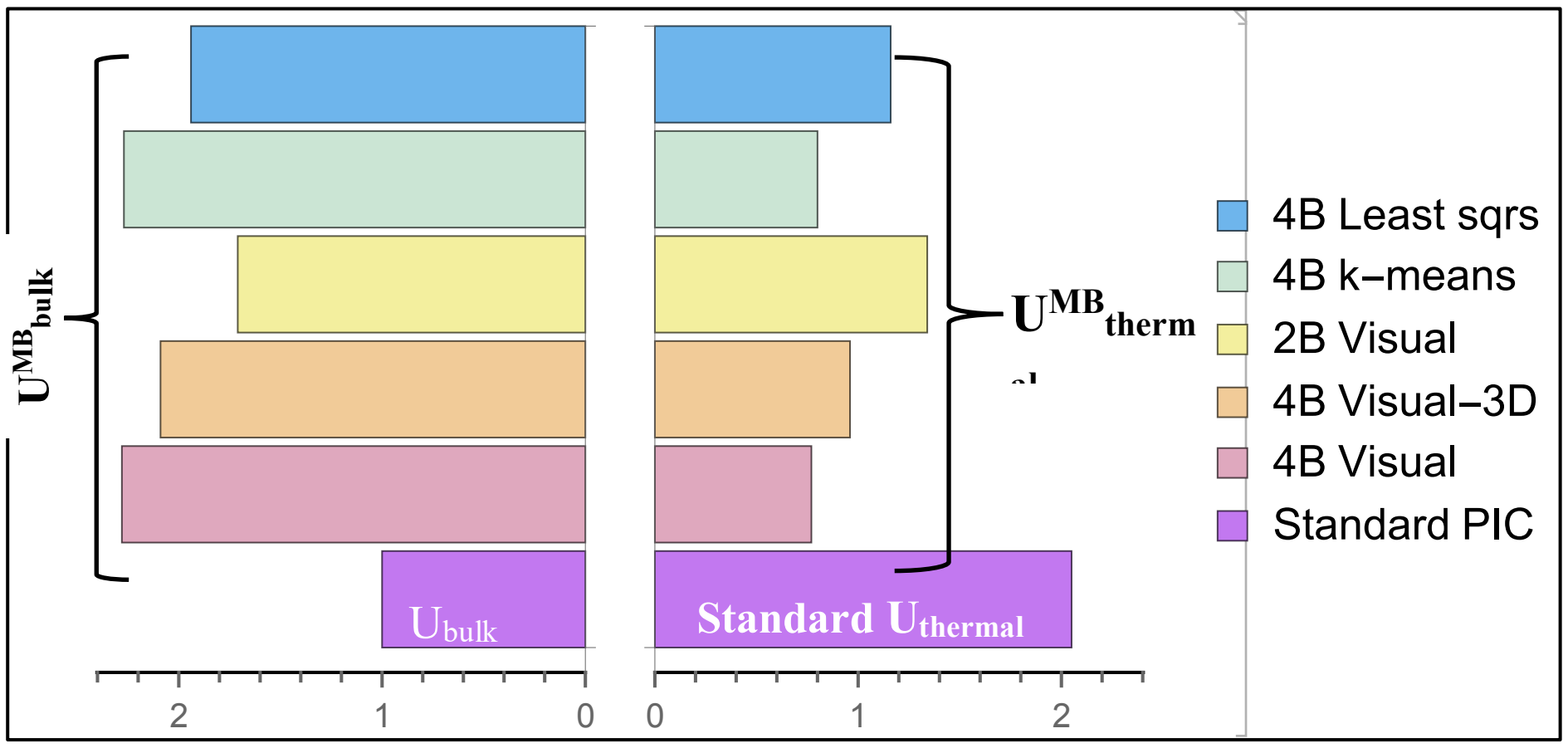

**Figure 6: Comparison of PIC ion energy density moments:** Purple: *Standard* bulk and thermal energy density moments from PIC velocity distribution, f(**v**); Five colors above purple: *multibeam* bulk (left) and thermal (right) energy density moments found by visual, k-means and least squares methods. Among these three are different examples of the visual method: the four-beam and two-beam cases based on 2D planar reduced distribution visualization in the Appendix and a four-beam case based on the 3D visualization in Fig. (5a). All (dimensionless) energies are in units of standard bulk energy from PIC data.

Next, we summarize our results for the standard and multi-beam energy *flux vectors* associated with the PIC f(**v**). Fig, (7) shows the *thermal* energy flux vector found by using the different methods. Recall that the thermal energy flux vector is the sum of the enthalpy flux vector and the heat flux vector, which is much smaller in magnitude than the enthalpy flux for this PIC f(**v**).

The black vector is the *standard* thermal energy flux moment, $\mathbf{Q}_{therm}$, from the PIC simulation, f(**v**). It is larger in magnitude than the other vectors, whch are all *multi-beam* thermal flux moment vectors. The multibeam vectors, $\mathbf{Q}^{MB}_{therm}$ are bunched together fairly close in angle to the standard $\mathbf{Q}_{therm}$. However, their magnitudes cover a range of values. The 2-beam visual method $\mathbf{Q}^{2B}_{therm}$ is the largest in magnitude of the multi-beam $\mathbf{Q}^{MB}_{therm}$ and the 4-beam k-means method $\mathbf{Q}^{4B}_{therm}$ is the smallest in magnitude. This ordering is consistent with the ordering of the thermal *energy densities* in Fig. (6), in which the PIC thermal energy density, $U^{2B}_{therm}$, found visually with *two* beams, is the largest of the $U^{MB}_{therm}$ and the $U^{4B}_{therm}$ found from k-means is the smallest. The pseudothermal energy flux is given by $\mathbf{Q}_{therm}$ - $\mathbf{Q}^{MB}_{therm}$. The large magnitude of the 2-beam thermal flux, $\mathbf{Q}^{2B}_{therm}$ leads to the smallest pseudothermal energy flux contained in the standard thermal moment, $\mathbf{Q}_{therm}$.

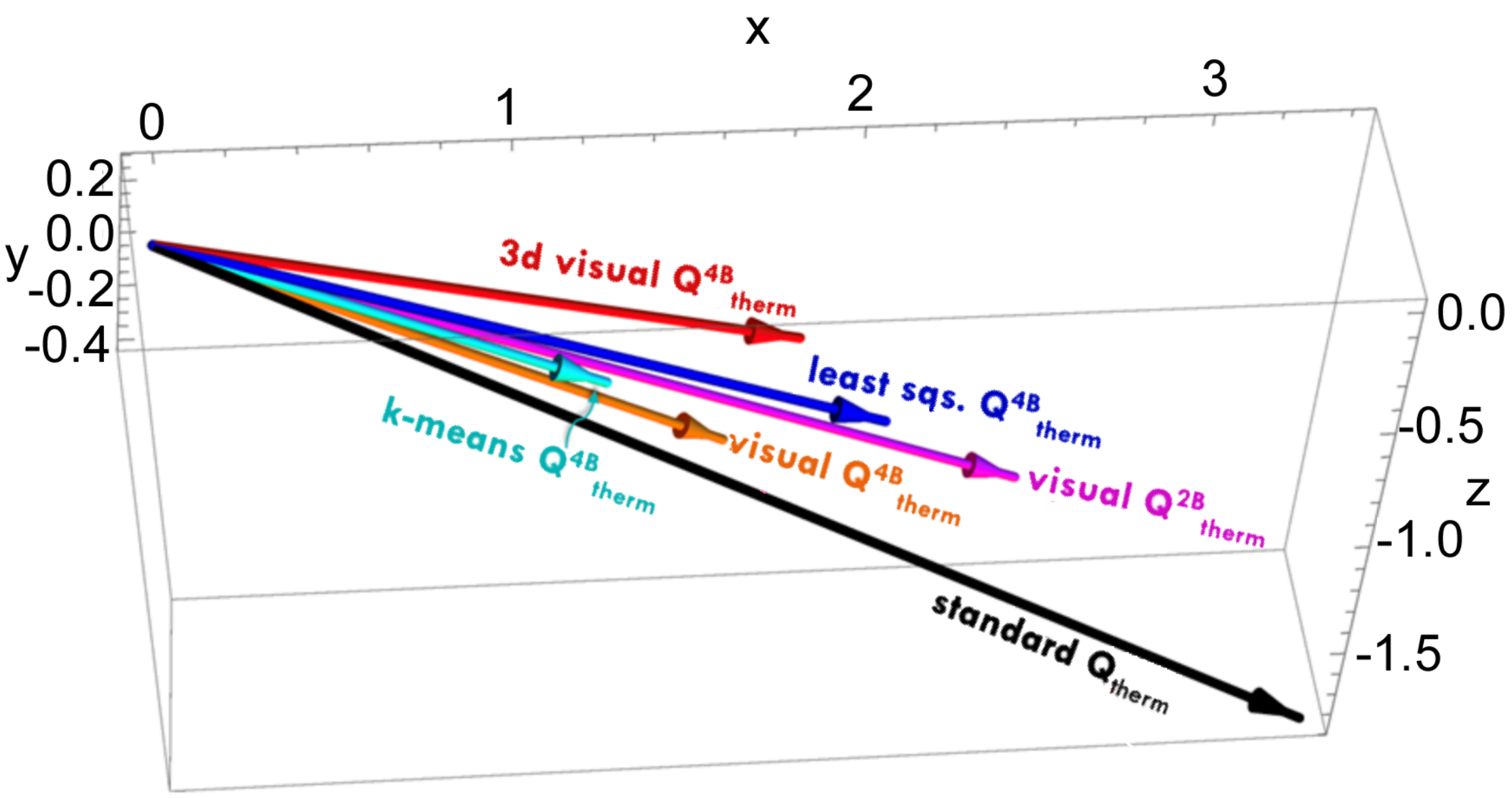

**Figure 7:** Comparison of standard and multi-beam *thermal* energy flux vectors for PIC f(**v**). Standard $\mathbf{Q}_{therm}$ is black. Three multi-beam thermal energy flux vectors, $\mathbf{Q}^{MB}_{therm}$, found by the visual method are displayed in warm colors, orange, magenta and red, representing, respectively, N = 4 beams and N = 2 beams, from reduced distributions and N = 4 beams from PIC 3D distribution in Fig. (3a). The vector moment, $\mathbf{Q}^{4B}_{therm}$, found from k-means is colored cyan and the $\mathbf{Q}^{4B}_{therm}$ found from least squares is colored blue.

The sum of thermal and bulk energy density fluxes of the PIC f(**v**) are shown in the vector diagrams of Fig. (8). The thick arrows show the sum of standard energy flux vectors: the bulk kinetic flux (thick blue) plus the thermal energy density flux (thick red). The *heat flux* is negligible compared to the *enthalpy flux* in all these thermal energy flux vectors.

The thin arrows are all four-beam energy fluxes. The thin red, green and tan arrows are the *thermal* energy flux vectors found, respectively, by the four-beam visual, k-means and least-squares methods. In comparison to the standard thermal energy flux vector they are all smaller in magnitude and make a small angle to it. Each is vectorially added to a different thin blue vector (multi-beam bulk K.E. flux) to equal an undecomposed energy density flux vector (not shown explicitly). These undecomposed energy density flux vectors are almost all the same length, although for the least squares method there is a noticeable difference (gap). Their magnitudes may be found in the Tables.

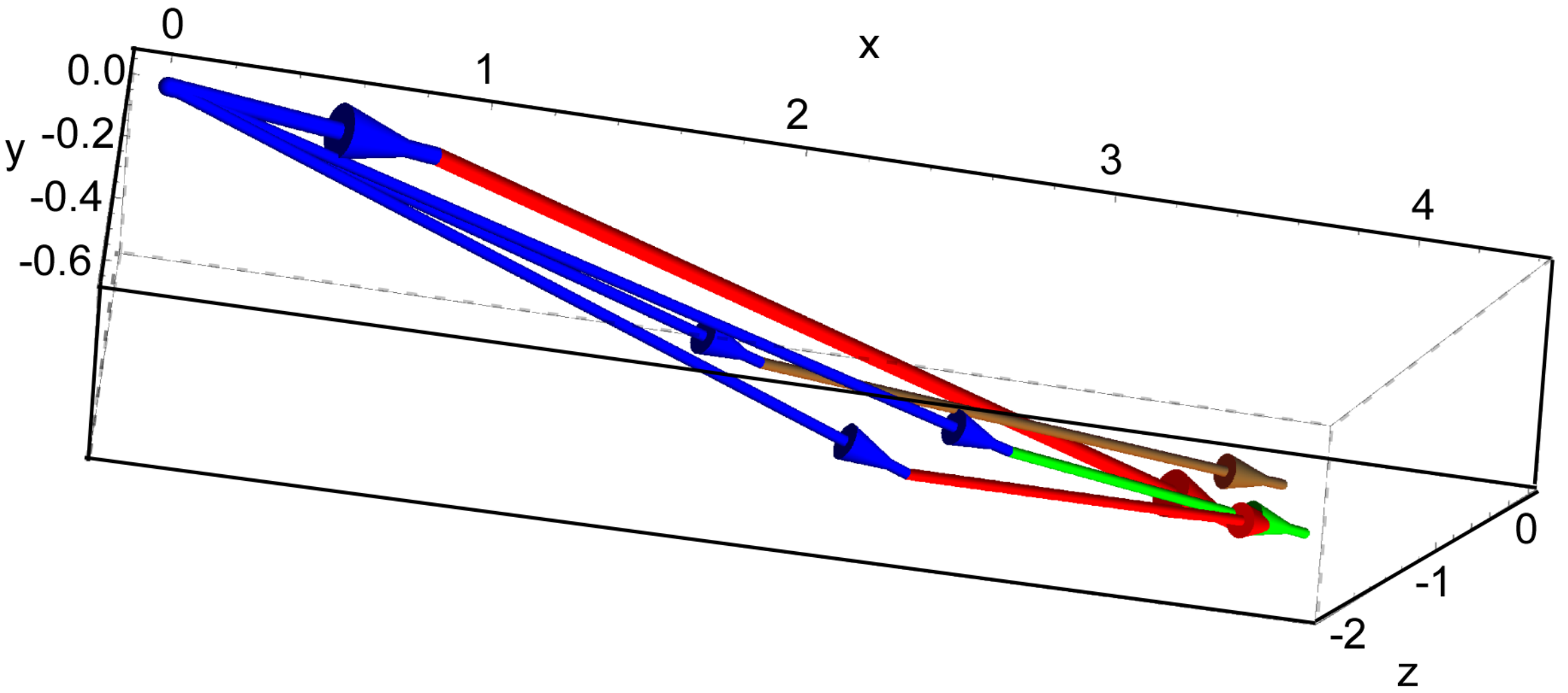

**Figure 8:** ***Multibeam*** **total energy flux moment vectors (thin arrows) compared to** ***standard*** **total energy flux moment vectors** for tail PIC simulation. The flux vectors have all been made dimensionless by dividing by the magnitude of the PIC bulk energy flux vector, $mnu^3/2$. The thick blue vector is the standard *PIC bulk energy flux*, of magnitude one. The standard PIC enthalpy flux is the thick red vector, and the much smaller PIC heat flux is not visible. The thin blue arrows are the multibeam bulk energy flux vectors found by the three multibeam methods. Each thin blue vector is joined to a different-colored thin vector, which is the multibeam *thermal energy flux moment* (mostly *enthalpy flux*), found by a different method: visual method indicated by thin red vector; k-means method indicated by thin green vector and least-squares method indicated by thin tan vector.

## 6. Moments of ion distribution, f(v) measured by MMS-4 during dayside reconnection

In this Section, four-beam moments are calculated for the velocity distribution, f(**v**), measured by the MMS Fast Plasma Instrument in the dayside magnetosphere during the MMS-4 reconnection event of October 2015, at 13:07:06 (Burch, et al, 2015). We will refer to it as the MMS f(v). This distribution and its moments differ in significant ways from the PIC simulation f(**v**) treated above in Section (5).

The multi-beam moments calculated by the *visual method* will be based on inspection of the "box"-processed ion velocity distribution, $f_{box}$(v), derived from the FPI skymap (Sec. 4.3.2). The multi-beam moments will be *compared* both to the standard moments of $f_{box}$(v) *and* to the FPI standard moments file. By contrast, multi-beam moments calculated by the k-means method are based on the ensemble of equally-weighted *particles* derived from $f_{box}$(v) (Sect. 4.3.3) and will be compared to the standard moments of $f_{box}$(v). Finally, the multi-beam moments calculated by the least-squares method will be based on the least-squares *fitted* ion velocity distribution, $f_{ls}$(v) and compared to the standard moments of both $f_{ls}$(**v**) and $f_{box}$(v). The ion velocity distributions appropriate for each method are summarized in Table (9)

**Table 9**: Ion velocity distribution used by each method to find multi-beam moments and standard moments for comparison. Dayside reconnection MMS4 Oct. 2015, 13:07:06 (Burch, et al, 2015).

| Method for finding multi-beam moments | Distribution used to find multi-beam moments | Compare multi-beam moments with |
|---|---|---|
| *Visual method* | $f_{box}$(**v**) | • FPI standard moments file<br>• standard moments of $f_{box}$(**v**) |
| *k-means method* | *Particle* distribution derived from $f_{box}$(**v**) | • standard moments of $f_{box}$(**v**) |
| *Least-squares method* | Fitted ion distribution, $f_{ls}$(**v**) | • standard moments of $f_{ls}$(**v**)<br>• standard moments of $f_{box}$(**v**) |

Once again *3D fly-around surface-contour plots* will be employed to visualize both f(**v**) and the effective ion "*beams"* corresponding to each of the three different methods for finding multi-beam moments**.**

*Reduced 2D velocity distributions* are used below (in Sect. 6.1) to visualize the MMS $f_{box}$(v) and to compare it to the PIC $f_{box}$(v).

*Field-aligned velocity coordinates* are used in both visualizations, with $v_{||}$ parallel to B, $v_{\perp 1}$, is in the direction of z x B, where z is the GSE coordinate perpendicular to the plane of the ecliptic and $v_{\perp 2}$ is in the direction of B x (z x B).

### 6.1 **REDUCED** ION VELOCITY DISTRIBUTION, $f_{box}(v)$, FROM MMS-4

The reduced ion velocity distributions of $f_{box}(v)$ derived from the FPI Skymap on MMS-4 during reconnection at the magnetopause Oct. 2015, at 13:07:06 (Burch, et al, 2015) are shown in Fig. (9) in terms of field-aligned coordinates, $\{v_{\|}, v_{\perp 1}, v_{\perp 2}\}$, and reduced velocity distributions, $F_a$, $F_b$, $F_c$, defined as

(10a) $F_a(v_{\|}, v_{\perp 1}) \equiv \int_{-\infty}^{\infty} dv_{\perp 2} f(v_{\|}, v_{\perp 1}, v_{\perp 2})$

(10b) $F_b(v_{\|}, v_{\perp 2}) \equiv \int_{-\infty}^{\infty} dv_{\perp 1} f(v_{\|}, v_{\perp 1}, v_{\perp 2})$

(10c) $F_c(v_{\perp 1}, v_{\perp 2}) \equiv \int_{-\infty}^{\infty} dv_{\|} f(v_{\|}, v_{\perp 1}, v_{\perp 2})$

Fig. (9) displays a *logarithmic* contour plot of the reduced distributions of the MMS ion $f_{box}(v)$ in field -aligned velocity coordinates. The standard moment flow velocity is indicated by the white plus signs. The red and black regions indicate that the MMS-measured ion distribution, f(v), is multi-beamed. A high-velocity ion cloud is seen to extend out beyond 500 km/s, although the magnitude of the MMS ion f(v) at such high velocities is down by at least five orders of magnitude. The high-velocity cloud is an important feature, not present in the PIC ion f(v) (Fig. 2). By contrast, the PIC f(v) falls to zero at around 500 km/s.

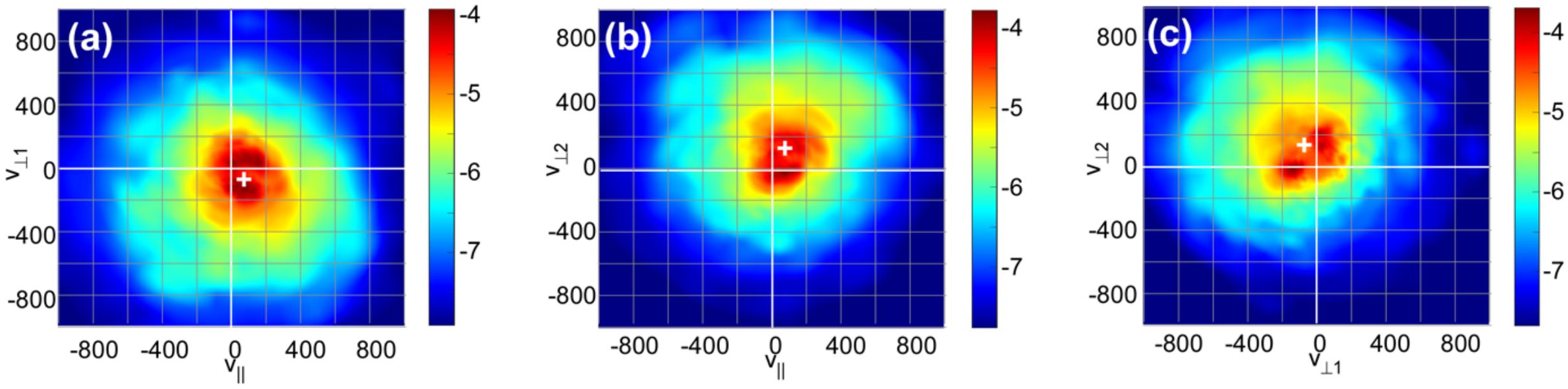

**Figure 9: Log of MMS reduced ion velocity distributions defined in Eqns. 10**. Velocities are in units of km/s. Note the high velocity "cloud" in the MMS f(**v**). It is not present in the PIC f(**v**) reduced ion velocity distributions shown in Fig. (1). The white plus marks are the projections of the standard flow velocity on the $v_{\|}$-$v_{\perp 1}$, $v_{\|}$-$v_{\perp 2}$ and $v_{\perp 1}$-$v_{\perp 2}$ planes.

### 6.2 STANDARD MOMENTS OF MMS ION VELOCITY DISTRIBUTIONS

Table (9) indicated the three *different kinds* of standard moments required to compare with the multi-beam moments found by the three different methods for the dayside reconnection event measured onboard and displayed in Fig. (9). These standard moments are given below in Tables 10-12

**Table 10: Standard moments from FPI moment file** during magnetopause reconnection Mean flow velocity, u, in km/s. Dimensionless thermal energy, $U_{thermal}$ in units of $U_{bulk} = Mnu^2/2$, pressure in units of $mnu^2$ and fluxes in units of $|Q_{bulk}| = Mnu^3/2$. Vector and matrix components are in field-aligned coordinate system in which parallel direction is along B and Perp1, Perp2 are in directions **z** x **B** and **B** x (**z** x **B**)

| $n_i Mu^2 = 3.62 \cdot 10^{-10}\ J/m^3$<br>$n_i Mu^3 = 0.0596\ mW/m^2$, n = 8.15 $cm^{-3}$<br>from **FPI standard** moments file | ⊥1 | ⊥2 | \|\| | Norm, Trace or Eigenvalues |
|---|---|---|---|---|
| Mean flow, **u** | -72.95 | 127.14 | 74.31 | u = \|**u**\| = 164 km/s |
| Pressure $P_{\perp 1,i}$ | 1.710 | 0.1661 | -0.0961 | Eigenvalues(**P**) = {1.902, 1.653, 1.443} |
| Pressure $P_{\perp 2,i}$ | 0.1661 | 1.755 | 0.0482 | |
| Presure $P_{\|\|,i}$ | -0.0961 | 0.0482 | 1.533 | Tr**P**= $U_{thermal}$ = *4.998* |
| $\mathbf{Q}_{bulk}$ | -0.444 | 0.774 | 0.452 | $\|\mathbf{Q}_{bulk}\|$ =1.00 |
| $\mathbf{Q}_{enthalpy}$ | -3.56 | 6.47 | 3.79 | $\|\mathbf{Q}_{enthalpy}\|$ =8.31 |
| $\mathbf{Q}_{heatflux}$ | -1.04 | 2.17 | -0.804 | $\|\mathbf{Q}_{heatflux}\|$ =2.54 |
| $\mathbf{Q}_{thermal} = \mathbf{Q}_{enthalpy} + \mathbf{Q}_{heatflux}$ | -4.60 | 8.64 | 2.99 | $\|\mathbf{Q}_{thermal}\|$ =10.24 |
| $\mathbf{Q} = \mathbf{Q}_{bulk} + \mathbf{Q}_{thermal}$ | -5.05 | 9.42 | 3.44 | $\|\mathbf{Q}\|$ = 11.23 |

**Table 11:** Standard **moments of** $f_{box}$(**v**) **(processed FPI)**.

| **Standard** moments of $f_{box}$(**v**)-processed FPI skymap | ⊥1 | ⊥2 | \|\| | Norm, Trace or Eigenvalues |
|---|---|---|---|---|
| Mean flow, **u** | -76.87 | 119.41 | 74.50 | u = \|**u**\| = 160 km/s |
| Pressure $P_{\perp 1,i}$ | 1.645 | 0.089 | -0.103 | Eigenvalues(**P**) = |
| Pressure $P_{\perp 2,i}$ | 0.089 | 1.747 | 0.128 | {1.809, 1.666, 1.289} |
| Presure $P_{\|\|,i}$ | -0.103 | 0.128 | 1.373 | Tr**P** = 4.765 = $U_{thermal}$ |
| $\mathbf{Q}_{bulk}$ | -0.468 | 0.727 | 0.453 | $\|\mathbf{Q}_{bulk}\|$ =0.976 |
| $\mathbf{Q}_{enthalpy}$ | -3.73 | 6.03 | 3.69 | $\|\mathbf{Q}_{enthalpy}\|$ = 7.99 |
| $\mathbf{Q}_{heatflux}$ | -1.79 | 2.50 | -1.30 | $\|\mathbf{Q}_{heatflux}\|$ = 3.34 |
| $\mathbf{Q}_{thermal} = \mathbf{Q}_{enthalpy} + \mathbf{Q}_{heatflux}$ | -5.52 | 8.53 | 2.39 | $\|\mathbf{Q}_{thermal}\|$ = 10.44 |
| $\mathbf{Q} = \mathbf{Q}_{bulk} + \mathbf{Q}_{thermal}$ | -5.99 | 9.26 | 2.84 | $\|\mathbf{Q}\|$ = 11.39 |

**Table 12 *Standard* moments of velocity distribution, $f_{fit}$(v) which is least squares fit to MMS $f_{box}$(v:** Mean flow velocity, u, in km/ss. Pressure in units of $mnu_n^2$ and fluxes in units of $|Q_{bulk}| = mnu_n^3/2$, from FPI moments file (Table 10). Vector and matrix components in field-aligned coordinate system in which the parallel direction is along **B, Perp1** is in direction **z x B** , and **Perp2** is in direction **B x (z x B)**

| **Standard** moments of $f_{ls}$(**v**), the least squares fit to MMS $f_{box}$(**v**) | ⊥1 | ⊥2 | ‖ | Norm, Trace or Eigenvalues |
|---|---|---|---|---|
| Mean flow, **u** | -62.21 | 112.35 | 74.67 | u = \|**u**\| = 148.6 km/s |
| Pressure $P_{\perp 1,i}$ | 1.640 | 0.083 | -0.120 | Eigenvalues(**P**) = |
| Pressure $P_{\perp 2,i}$ | 0.083 | 1.893 | 0.145 | {1.943, 1.693, 1.326} |
| Presure $P_{\|\|,i}$ | -0.120 | 0.145 | 1.429 | Tr**P** = $U_{thermal}$ = 4.962 |
| $\mathbf{Q}_{bulk}$ | -0.309 | 0.559 | 0.371 | $\|\mathbf{Q}_{bulk}\|$ =0.739 |
| $\mathbf{Q}_{enthalpy}$ | -3.116 | 6.050 | 3.842 | $\|\mathbf{Q}_{enthalpy}\|$ = 7.814 |
| $\mathbf{Q}_{heatflux}$ | -1.613 | 2.268 | -1.328 | $\|\mathbf{Q}_{heatflux}\|$ = 3.083 |
| $\mathbf{Q}_{thermal} = \mathbf{Q}_{enthalpy} + \mathbf{Q}_{heatflux}$ | -4.728 | 8.318 | 2.514 | $\|\mathbf{Q}_{thermal}\|$ = 9.892 |
| $\mathbf{Q} = \mathbf{Q}_{bulk} + \mathbf{Q}_{thermal}$ | -5.038 | 8.876 | 2.885 | $\|\mathbf{Q}\|$ = 10.606 |

All three sets of *standard* moments are in dimensionless units based on the standard $U_{bulk}$ and $|\mathbf{Q}_{bulk}|$ from the FPI moments file in Table 10. The thermal energy densities, $U_{thermal}$ from Tables 10, 11 and 12, respectively are 4.99, 4.77, and 4.96. The magnitudes of the undecomposed energy density flux vectors, |**Q**|, are, respectively, 11.23, 11.39, and 10.61. The three different *standard* moments have flow velocities, 164, 160, and 148.6 km/s.

Fig. (10) shows the three standard moments, $\mathbf{Q}_{bulk}$, $\mathbf{Q}_{enthalpy}$ and $\mathbf{Q}_{heatflux}$ from the FPI standards moments file, for $f_{box}$ and for $f_{ls}$. Note that the total moment, $\mathbf{Q} = \mathbf{Q}_{bulk} + \mathbf{Q}_{enthalpy} + \mathbf{Q}_{heatflux}$ is different for each of the three distribution functions.

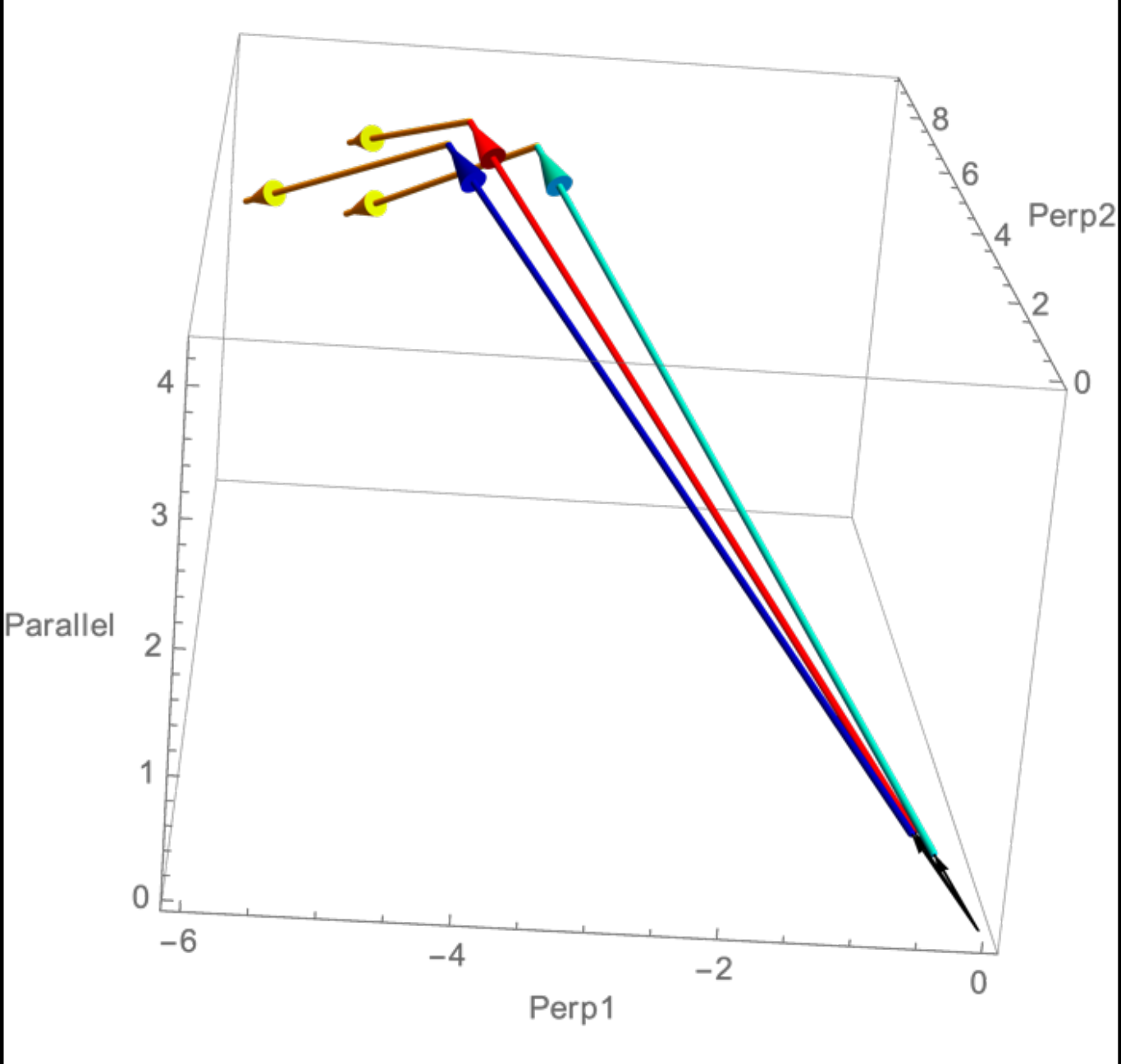

**Figure 10: Standard moments, $\mathbf{Q}_{bulk}$, $\mathbf{Q}_{enthalpy}$** and $\mathbf{Q}_{heatflux}$ from FPI standard moments file (thin black, red, yellow), for $f_{box}$ (thin black, blue, yellow), and for $f_{ls}$ (thin black cyan, yellow).

6.3 MMS f(**v**) MULTI-BEAM MOMENTS

Fig. 11 shows 3-D velocity contour maps of the *MMS ion velocity distribution*, f(**v**), whose reduced distributions are displayed in Fig. (9). Fig 11a shows a surface contour map of the MMS $f_{box}(\mathbf{v})$ with the standard flow velocity denoted by a black plus-mark and four visually estimated beam velocity centroids marked with colored plus marks and connected by a tetrahedron. The standard and visual multi-beam pressure ellipsoids are displayed in pink and black, respectively. They are closer together in this MMS case than in the PIC analysis (Fig. 4b). "Figs. 11c and 11d show the least squares fit to $f_{box}(\mathbf{v})$ and the four beams and beam centroids, where one of the four is the composite 3-kappa cloud**.**

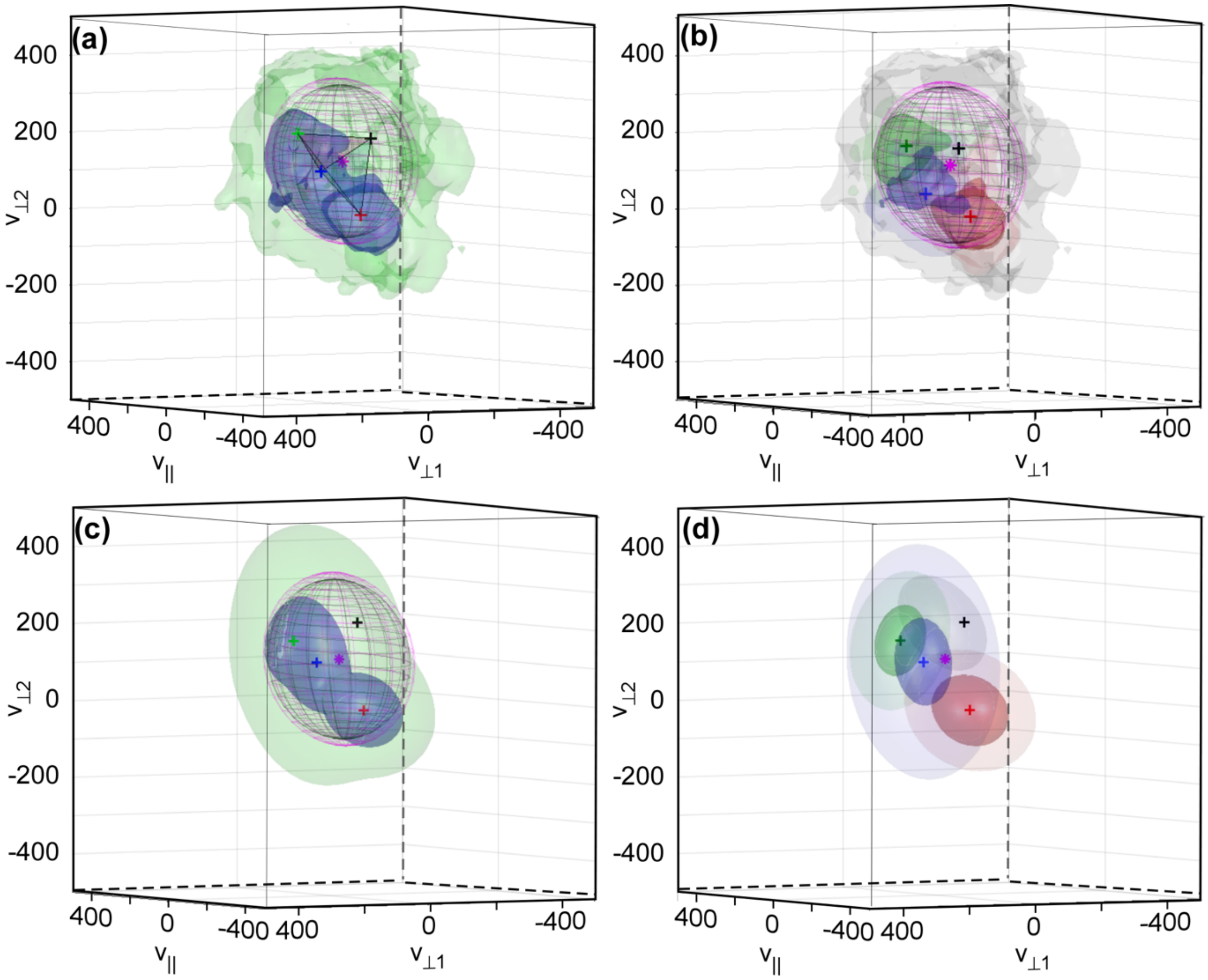

**Figure 11: 3D visualizations of MMS f(v) for each of the three methods, showing beams, centroid velocities and pressure ellipsoids (**standard in pink and multibeam in black**.** (a): Surface contour map of MMS $f_{box}(v)$. Inner surfaces are at 10% of the global maximum of f(v) and outer surfaces are at 1%. Colored plus-marks indicate four visually estimated beam velocities connected by tetrahedron with magenta star inside indicating standard flow velocity **u** . (b) shows k-means cluster decomposition (with pedestal). (c,d) Nonlinear least squares fit to $f_{box}(v)$. The sum of components is plotted in (a) the individual components are color coded in (d)

Multi-beam moments found by each method are organized below in Tables (13-15) for comparison with each other and with the standard moments given in Tables (10-12). In Section (6.4) we compare multi-beam and standard *energy and energy flux moments* found by the different methods.

*Visual method*

The visual method uses the FPI standard moments file. From Table (13), below, it is seen to yield a 4-beam thermal energy density, $U^{4B}_{therm}$= 4.42*,* which is less than the FPI moments file standard thermal energy density, $U_{therm}$ = 5.02. The difference of ΔU = 0.60, means that about 12% of the thermal energy in the FPI moments file is pseudo*thermal energy.* This is a *small* correction to $U_{therm}$ compared to the PIC pseudothermal energies found in Section 5. If the 4-beam energy density $U^{4B}_{therm}$ is compared to the moments of $f_{box}$ instead of to the moments in the FPI moments file, the pseudothermal energy is even greater (see Table (9))

**Table 13. Four-beam moments of MMS f(v) from *visual* method** displayed in Fig. (11a). Beam centroid velocities are in km/sec. The four colors refer to the colors of the plus marks in Fig. 11a. Pressure is in units of $mnu_n^2$ and fluxesare in units of $|Q_{bulk}| = mnu_n^3/2$, from the FPI moments file (Table 10).

| ***4-beam moments from visual method based on FPI f(v)*** | **⊥1 comp.** | **⊥2 comp.** | **\|\| comp.** | ***Related scalars*** | |
|---|---|---|---|---|---|
| *Beam 1 velocity (black)* | -155 | 185 | 70 | η = 0.352 | \|**v**\| = 251.3 |
| *Beam 2 velocity (red)* | -125 | -15 | 70 | η = 0.252 | \|**v**\| = 144.0 |
| *Beam 3 velocity (green)* | 35 | 200 | 120 | η = 0.255 | \|**v**\| = 235.9 |
| *Beam 4 velocity (blue)* | 30 | 105 | 10 | η =0.141 | \|**v**\| = 109.7 |
| $P^{MB}_{\perp 1,i}$ | 1.432 | 0.097 | -0.115 | *Eigenvalues of* $P^{MB}$= {1.606, 1.485, 1.302} | |
| $P^{MB}_{\perp 2,i}$ | 0.097 | 1.470 | 0.007 | | |
| $P^{MB}_{\parallel,i}$ | -0.115 | 0.007 | 1.491 | $TrP^{MB}= U^{MB}_{therm}$=4.393 | |
| $\mathbf{Q^{MB}_{bulk}}$ | -0.801 | 1.589 | 0.821 | $\|Q^{MB}_{bulk}\|$= 1.960 | |
| $\mathbf{Q^{MB}_{therm} = Q^{MB}_{enth} + Q^{MB}_{htflux}}$ | -4.243 | 7.825 | 2.621 | $\|Q^{MB}_{therm}\|$ =9.279 | |
| $\mathbf{Q^{MB} = Q^{MB}_{bulk} + Q^{MB}_{therm}}$ | -5.044 | 9.414 | 3.442 | $\|Q^{MB}\|$ = 11.221 | |

The total 4-beam energy flux found by the visual method is $|\mathbf{Q}^{MB}|$ = 11.22, which is identical (by construction) to the total standard energy density flux, |**Q**|, from the FPI moment file (Table 10). Note that the magnitude of the bulk energy flux vector, $|\mathbf{Q}^{MB}_{bulk}|$**,** is much less than the magnitude of the thermal energy flux vector, $|\mathbf{Q}^{MB}_{therm}|$**.**

*k-means with a 1% pedestal*

Table 14 gives multi-beam moments found by the k-means method with k = 3 and a 1% pedestal (Sect 4.4). For k-means and the least square method,both $\mathbf{Q}^{MB}_{htflux}$ and $\mathbf{Q}^{MB}_{enthalpy}$ can be found separately, whereas for the visual method only the sum, $\mathbf{Q}^{MB}_{therm}$ can be found. The magnitude, $|\mathbf{Q}^{MB}_{htflux}|$, is 1.72, which is about half the magnitude, 3.34, of the standard $|\mathbf{Q}_{htflux}|$ (Table 11). Note, also, that the k-means *energy density* of 4.390 is virtually identical to the value, 4.393, found by the visual method (Table 13). However, the magnitude $|\mathbf{Q}^{MB}_{bulk}|$ of 1.444 found by k-means is less than the value, 1.960, found by the visual method. The total multibeam energy flux magnitude, $|\mathbf{Q}^{MB}|$ is 11.32 for k-means, compared to 11.22 for the visual method.

**Table 14:** Multi-beam moments of MMS $f_{box}(\mathbf{v})$ from **k-means method** with three beams (k =3) and a 1% pedestal (Beam 1). Velocities in km/s. Pressure in units of $mnu_n^2$. **Q**'s in units of $|Q_{bulk}| = mnu_n^3/2$.

| ***k-means four-beam moments of MMS f(v)*** | **⊥1 comp.** | **⊥2 comp.** | **\|\| comp.** | ***Related scalar/vector*** | |
|---|---|---|---|---|---|
| *Beam 1 (pedestal) velocity* | -100.8 | 162.4 | 72.0 | $\eta$ =0.576 | \|v\| = 204 |
| *Beam 2 velocity* | -136.8 | -18.6 | 74.2 | $\eta$ = 0.188 | \|v\| = 157 |
| *Beam 3 velocity* | 27.2 | 169.7 | 121.8 | $\eta$ = 0.146 | \|v\| = 211 |
| *Beam 4 velocity* | 33.3 | 49.8 | 12.3 | $\eta$ = 0.090 | \|v\| = 61.2 |
| $P^{MB}_{\perp 1,i}$ | 1.506 | 0.052 | -0.107 | *Eigenvalues of* $\mathbf{P}^{MB}$ = {1.587, 1.558, 1.245} | |
| $P^{MB}_{\perp 2,i}$ | 0.052 | 1.543 | 0.102 | | |
| $P^{MB}_{\|\|,i}$ | -0.107 | 0.102 | 1.340 | $Tr\mathbf{P}^{MB} = U^{MB}_{therm} = 4.390$ | |
| $\mathbf{Q}^{MB}_{bulk}$ | -0.648 | 1.115 | 0.649 | $\|\mathbf{Q}^{MB}_{bulk}\| = 1.444$ | |
| $\mathbf{Q}^{MB}_{therm} = \mathbf{Q}^{MB}_{enth} + \mathbf{Q}^{MB}_{htflux}$ | -5.308 | 8.086 | 2.172 | $\|\mathbf{Q}^{MB}_{therm}\| = 9.913$ | |
| $\mathbf{Q}^{MB}_{htflux}$ | -0.853 | 0.782 | -1.278 | $\|\mathbf{Q}^{MB}_{htflux}\| = 1.724$ | |
| $\mathbf{Q}^{MB} = \mathbf{Q}^{MB}_{bulk} + \mathbf{Q}^{MB}_{therm}$ | -5.956 | 9.201 | 2.821 | $\|\mathbf{Q}^{MB}\| = 11.318$ | |

*Modified least-squares fit to an MMS f(**v**) with a high-energy cloud.*

Next multi-beam moments of the MMS f(**v**) are found by the least-squares method.

- First the cloud, $f_{cloud}$, is modeled from the MMS f(**v**) by fitting the *log* of the distribution at high velocities to the *log* of the sum of three kappa distributions (beam 1, beam 2 and beam 3). The sum of these three distributions is identified as $f_{cloud}(\mathbf{v})$.
- Next, the cloud distribution is subtracted from f(v) and the difference $[f(v)-f_{cloud}(\mathbf{v})]$ is fit separately with a sum of three different kappa functions (beam 4, beam 5 and beam 6), which are identified as *core beams* to differentiate them from the *cloud*.
- Rather than adding together the standard moments of beams 1, 2, 3, 4, 5 and 6 to get (six-beam) multi-beam moments, a new strategy is employed to improve the cloud modeling:

- One standard moment of the *ensemble* of the three *cloud beams*, 1, 2, and 3 is taken and added to the three standard moments of *each* of the *core beams* 4, 5, and 6 to produce (four-beam) multi-beam moments of f(**v**).
- This has the advantage that the cloud contribution to the multi-beam moments of f(**v**) has a non-vanishing standard heat-flux moment, whereas there is no standard heat-flux for any of the single kappa-function beams, due to their symmetry.

The beam centroid velocities and partial densities from the fits are given explicitly in Table (15), and in the number-density flux-vector diagram in Fig. (12). Note that $n_c\mathbf{u}_c = n_1\mathbf{u}_1 + n_2\mathbf{u}_2 + n_3\mathbf{u}_3$ is the standard number density flux of the set of three beams, 1, 2, and 3 representing the cloud. Also, n**u** is the standard number density flux of the six beams representing the modified least squares fit to the MMS f(**v**).

**Table 15:** Centroid velocities and fractional densities of the beams in a (modified) least-squares fit to the MMS f(v). Beams 1, 2, and 3 belong to the cloud, the *cloud flow* is from the standard flux of the three cloud beams and beams 4, 5 and 6 are core beams.

| Least-squares fit to MMS f(**v**) | Mean flow, **u** | Beam1 $\mathbf{u}_1$ | Beam2 $\mathbf{u}_2$ | Beam3 $\mathbf{u}_3$ | *Cloud flow*, $\mathbf{u}_c$ | Beam4 $\mathbf{u}_4$ | Beam5 $\mathbf{u}_5$ | Beam6 $\mathbf{u}_6$ |
|---|---|---|---|---|---|---|---|---|
| ⊥1 | -62 | -139 | -111 | -124 | -130 | -136 | 33 | 35 |
| ⊥2 | 112 | 212 | 160 | 206 | 205 | -24 | 159 | 109 |
| ‖ | 75 | -44 | -12 | 212 | 94 | 73 | 146 | 23 |
| *NORM* | *149* | *258* | *195* | *321* | *260* | *156* | *219* | *116* |
| $\eta_j$ | 1.001 | 0.131 | 0.021 | 0.171 | 0.323 | 0.257 | 0.130 | 0.291 |

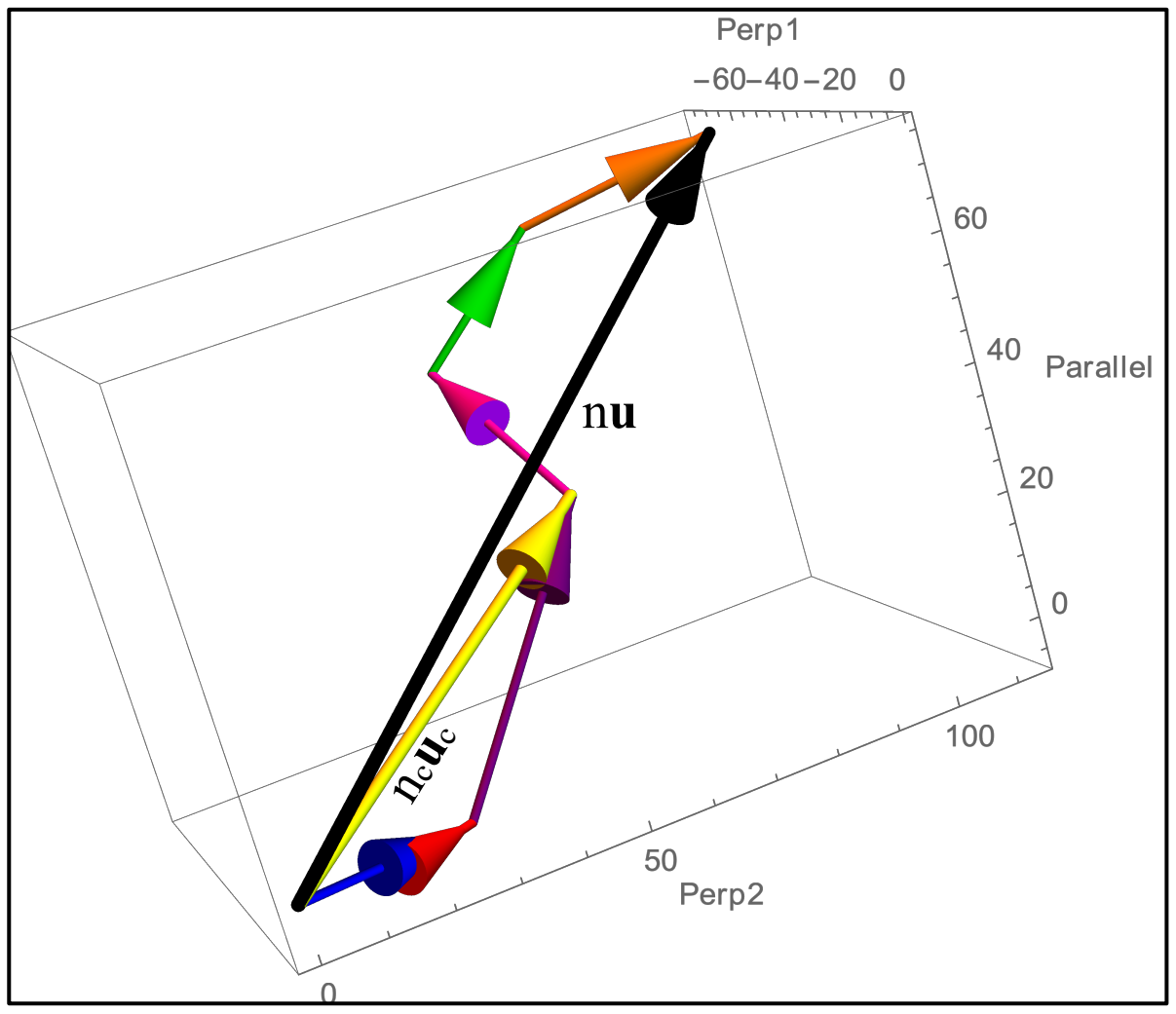

**Figure 12:** Density flux vectors for the six beams in Table 15 and standard moment flux vectors of cloud ($n_c\mathbf{u}_c$) and overall (n**u**).

Figs. (11c, d) shows the shapes of beams 4, 5, 6 and of the cloud (sum of beams 1, 2, 3) in this modified least-squares fit and their velocity centroids, $\mathbf{u}_4$, $\mathbf{u}_5$, $\mathbf{u}_6$ and $\mathbf{u}_c$, together with the multibeam and standard pressure ellipsoids.

The multi-beam (four-beam) moments of the modified least squares fit, $f_{fit}(\mathbf{v})$ to the MMS $f_{box}(\mathbf{v})$ are given in Table (16). The multi-beam energy density is 4.36, compared to the standard energy density of the least squares fitted distribution (Table 12), which is 4.96.

The magnitude of the heat-flux is 1.96, which is lower than the standard heat flux, 3.03, of the *fit distribution*, $f_{fit}$ (Table 12). The magnitude of the undecomposed energy flux, $|\mathbf{Q}^{MB}|$, is equal to the value of the standard undecomposed energy flux, $\mathbf{Q}$ = 10.606 (Table 12).

| ***Four-beam moments of least squares fit, $f_{fit}$, to MMS $f_{box}(\mathbf{v})$*** | ⊥1 comp. | ⊥2 comp. | \|\| comp. | ***Related scalar/vector*** |
|---|---|---|---|---|
| $P^{MB}{}_{\perp 1,i}$ | 1.390 | 0.045 | -0.084 | *Eigenvalues of $\mathbf{P}^{MB}$= {1.642, 1.459, 1.262}* |
| $P^{MB}{}_{\perp 2,i}$ | 0.045 | 1.602 | 0.104 | |
| $P^{MB}{}_{\|\|,i}$ | -0.084 | 0.104 | 1.371 | $Tr\mathbf{P}^{MB} = U^{MB}{}_{therm}$ = 4.362 |
| $Q^{MB}{}_{bulk}$ | -0.752 | 1.298 | 0.791 | $\|\mathbf{Q}^{MB}{}_{bulk}\|$ = 1.696 |
| $Q^{MB}{}_{enth}$ | -4.509 | 8.180 | 3.945 | $\|\mathbf{Q}^{MB}{}_{enth}\|$ = 10.146 |
| $Q^{MB}{}_{htflux}$ | -0.223 | -0.609 | -1.850 | $\|\mathbf{Q}^{MB}{}_{htflux}\|$ =1.960 |
| $Q^{MB}{}_{therm} = Q^{MB}{}_{enth} + Q^{MB}{}_{htflux}$ | -4.285 | 7.579 | 2.095 | $\|\mathbf{Q}^{MB}{}_{therm}\|$ = 8.955 |
| $Q^{MB} = Q^{MB}{}_{bulk} + Q^{MB}{}_{therm}$ | -5.038 | 8.876 | 2.885 | $\|\mathbf{Q}^{MB}\|$ = 10.606 |

**Table 16:** Multi-beam moments of modified *least-squares fit*, $f_{fit}(\mathbf{v})$ to MMS $f(\mathbf{v})$. Pressure in units of $mnu_n^2$ and fluxes in units of $|Q_{bulk}| = mnu_n^3/2$, from FPI moments file (Table 10).

## 6.4 COMPARISON OF MULTI-BEAM MOMENTS OF MMS f(v)

Here we summarize the results and significance of the three methods for finding multi-beam moments of the MMS ion velocity distribution, $f(\mathbf{v})$. There are three different *standard* moments: those from the MMS FPI Moments File (Table 10), those based on $f_{box}$ (Table 11) and those based on $f_{fit}$ (Table 12). The FPI Moments File is based on the FPI Skymap, which includes more higher energy ions than in $f_{box}$ or $f_{fit}$. Note that he higher energy ions must be arranged in velocity-space so as to make the magnitude of the standard heat flux from the FPI moments file (Table 10) *lower* than the magnitude of the standard heat flux for $f_{box}(\mathbf{v})$ (Table 11).

### *Energy density moments compared*

In Fig. (13) scalar multi-beam energy densities, $U^{MB}{}_{therm}$ and $U^{MB}{}_{bulk}$ found by the three methods are compared with each other and with the standard energy densities from the Moments File, from $f_{box}$ and from $f_{fit}$ (the least squares fit to $f_{box}$). Thermal energy densities are on the right, and bulk thermal energy densities are on the left. For each pair, the *undecomposed* U is the sum of $U^{MB}{}_{therm}$ and $U^{MB}{}_{bulk}$ (i.e., the sum of the lengths of the right and left bars of a given color). Once again, the pseudo-thermal energy density is $\Delta U = U_{thermal} - U^{MB}{}_{thermal} = U^{MB}{}_{bulk} - U_{bulk}$ for each method. Visually, ΔU is the difference in length between the multibeam and standard barsfor each method in Fig. 13. Comparing the energy densities here with those in Fig. 6 for the PIC $f(\mathbf{v})$, it is evident that the ΔU are larger for the PIC $f(\mathbf{v})$. This is because there are fewer high energy ions (i.e., not much of a high-energy cloud) in the PIC $f(\mathbf{v})$.

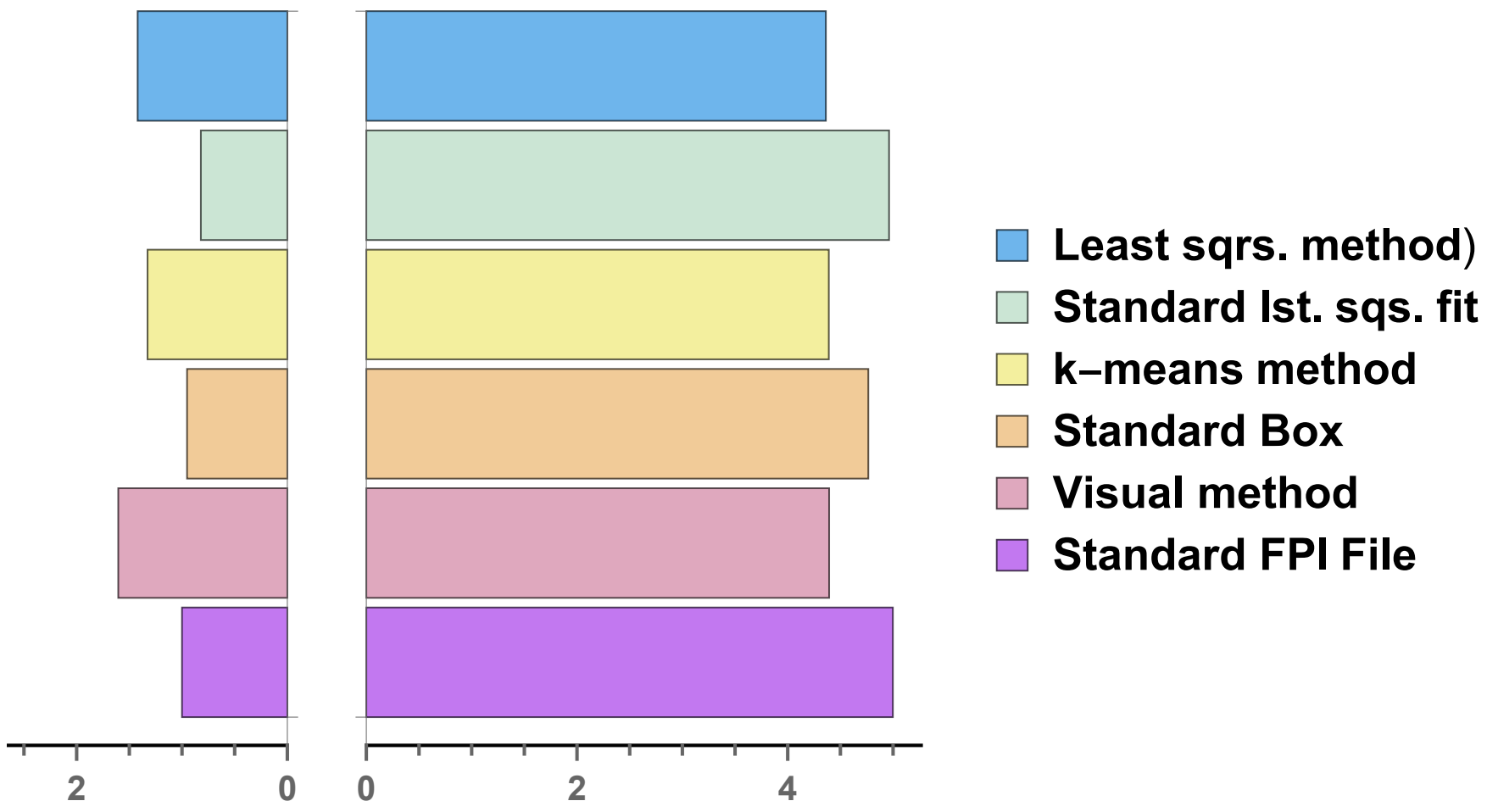

**Figure 13:** Standard and multibeam thermal (right bar) and bulk (left bar) energy densities. All (dimensionless) energies are in units of standard bulk energy from the FPI moments file.

Recall that for each method, the *total* standard energy density and the *total* multibeam energy density are equal, $U = U^{MB}$. Hence, for each method, the sum of left and right bar lengths are the same for both standard and multibeam moments. However this length differs slightly among the three methods, unlike the PIC case (Fig. 6), for which they were all essentially equal. Also note that $U^{MB}_{thermal} = U_{thermal} - \Delta U$, and $U^{MB}_{bulk} = U_{bulk} + \Delta U$, so that, for each method, the difference between standard and multibeam bars is the same at both left and right.

For the visual method, subtracting $U^{MB}_{thermal}$ from the standard $U_{thermal}$ using Tables (13) and (10) gives $\Delta U = 4.998 - \mathit{4.393} = 0.605$. This is 12% of $U_{thermal}$. Since the standard bulk moment, $U_{bulk}$, is 1 in our units, $U^{MB}_{bulk}$ becomes 1.6, a significant 60% correction to $U_{bulk}$.

The values of $U^{4B}_{thermal}$ found by the k-means and least squares methods are close to the value found by the 4-beam visual method.

*Energy density flux vector moments compared*

Next, we compare standard and multi-beam energy density *flux* moments of the MMS velocity distribution, f(**v**). These moments are vectors rather than scalars.

In Fig. (14) vector diagrams show the decomposition of the MMS multi-beam energy flux moments, $\mathbf{Q}^{MB}$ into a sum of a bulk, enthalpy and heat fluxes, $\mathbf{Q}^{MB}_{bulk}$, $\mathbf{Q}^{MB}_{enthalpy}$ and $\mathbf{Q}^{MB}_{heatflux}$ by the k-means and least squares methods and into a sum of $\mathbf{Q}^{MB}_{bulk}$ and $\mathbf{Q}^{MB}_{thermal} = \mathbf{Q}^{MB}_{enthalpy} + \mathbf{Q}^{MB}_{heatflux}$ by the visual method. Comparisons are made to the corresponding *standard* moments (Figs. 10, 11 and 12). Once again, $\mathbf{Q}^{MB} = \mathbf{Q}$, and the decomposition of $\mathbf{Q}^{MB}$ may be written for all methods in terms of $\mathbf{Q}^{MB}_{thermal} = \mathbf{Q}_{thermal} - \Delta\mathbf{Q}$ and $\mathbf{Q}^{MB}_{bulk} = \mathbf{Q}_{bulk} + \Delta\mathbf{Q}$, where the pseudo-thermal energy density flux vector is $\Delta\mathbf{Q} = \mathbf{Q}_{thermal} - \mathbf{Q}^{MB}_{thermal} = \mathbf{Q}^{MB}_{bulk} - \mathbf{Q}_{bulk}$ (Table 2).

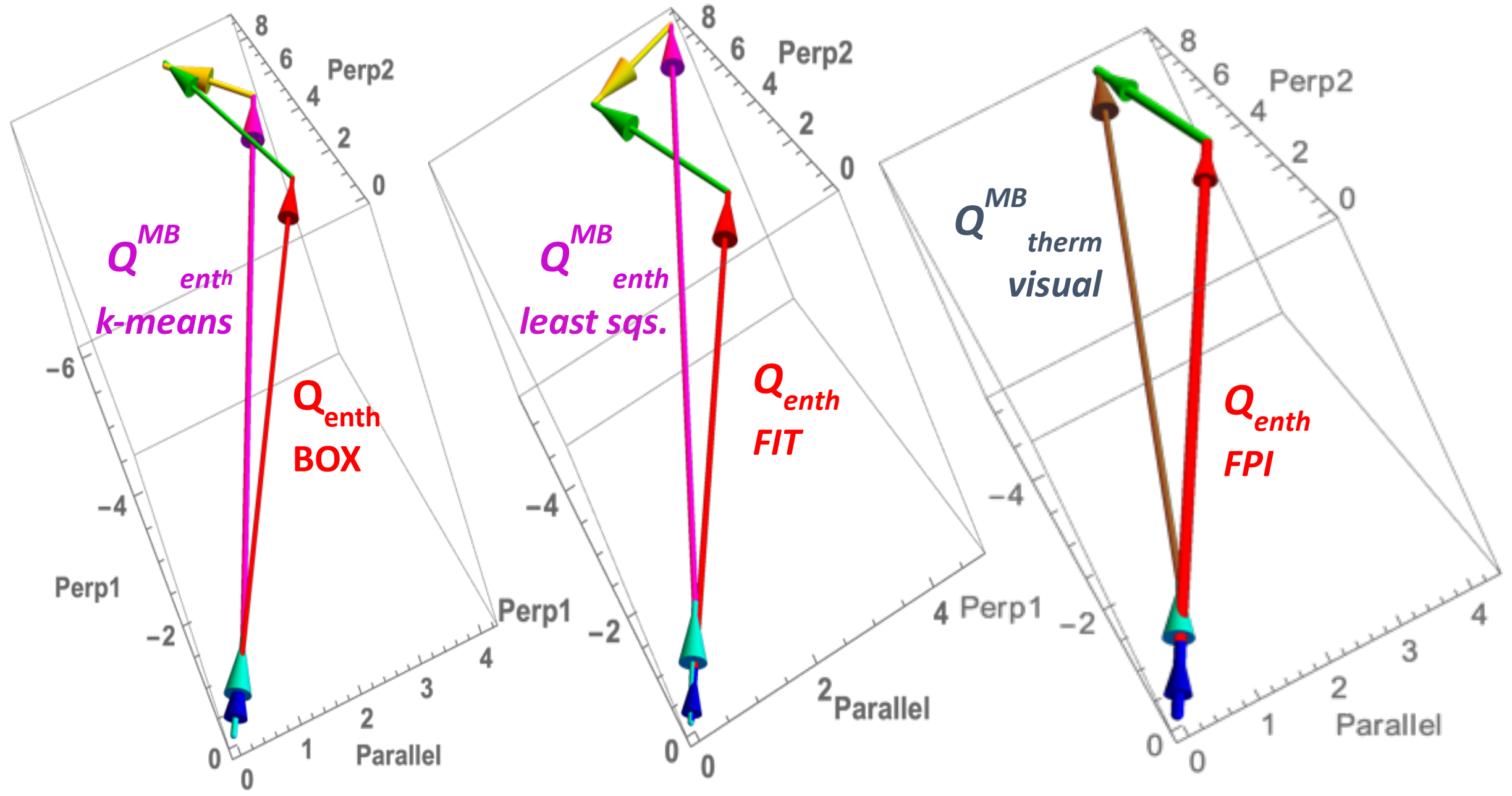

**Figure 14: Sums of multibeam and corresponding standard energy density fluxes (bulk, enthalpy, and heat flux) for each of the three methods.** Enthalpy fluxes are in red for standard and in magenta for multi-beam. Bulk kinetic energy fluxes are in blue for standard and in cyan for multi-beam. Heat fluxes are in green for standard and in yellow for multi-beam. For the visual method the thermal energy density flux is in brown.

The small blue and cyan vectors are the $\mathbf{Q}_{\text{bulk}}$ and $\mathbf{Q}^{\text{MB}}_{\text{bulk}}$ moments appropriate to each case. The red and green vectors are the standard moments $\mathbf{Q}_{\text{enthalpy}}$ and $\mathbf{Q}_{\text{heatflux}}$. For the k-means and least squares methods *(left* and *center figures)* the magenta and yellow vectors are the multi-beam moments, $\mathbf{Q}^{\text{MB}}_{\text{enthalpy}}$ and $\mathbf{Q}^{\text{MB}}_{\text{heatflux}}$. Note that the multi-beam enthalpy fluxes are longer than the standard enthalpy fluxes and that the multi-beam heat fluxes are shorter than and almost at right angles to the standard heat fluxes. The multi-beam heat fluxes are also significantly shorter than the corresponding multi-beam enthalpy fluxes in this, the MMS4 spacecraft frame of the measured f(**v**). This is also true of the magnitudes of the standard heat fluxes compared to those of the standard enthalpy fluxes, a result also found in other magnetospheric measurements (Eastwood, et al, 2013).

The standard and multi-beam enthalpy fluxes are *frame dependent*. The standard heat flux vector, $\mathbf{Q}_{\text{heatflux}}$, is independent of the standard flow velocity, **u**, and hence is *frame independent*. In the frame where $\mathbf{u} = 0$ (*rest frame),* $\mathbf{Q}_{\text{enthalpy}}$ vanishes (since it is proportional to **u)** so, in this frame, the standard thermal energy flux, $\mathbf{Q}_{\text{thermal}} = \mathbf{Q}_{\text{enthalpy}} + \mathbf{Q}_{\text{heatflux}}$, reduces to $\mathbf{Q}_{\text{heatflux}}$. The multi-beam heat flux is also independent of **u** because it is a sum of the standard heat fluxes of the N beams. However, since each of the beams generally has a different centroid velocity, there is no frame in which the sum of the standard enthalpy fluxes, $\mathbf{Q}^{\text{MB}}_{\text{enthalpy}}$, vanishes. Nevertheless, the ratio, $|\mathbf{Q}^{\text{MB}}_{\text{enthalpy}}|/|\mathbf{Q}^{\text{MB}}_{\text{heatflux}}|$ will be frame-dependent and the standard and multi-beam energy fluxes should be evaluated in special frames such as the frame of the x-point to get more meaningful physical results than those in Fig. 14.

The visual method (*shown at right*) does not yield a multi-beam heat flux moment, $\mathbf{Q}^{MB}_{heatflux}$, but only gives the multi-beam thermal flux, $\mathbf{Q}^{MB}_{thermal}$ (brown vector). The directions of the $\mathbf{Q}^{MB}_{thermal}$ vectors found by k-means and least squares can be surmised from the other images and all point in roughly the same direction. The respective magnitudes, $|\mathbf{Q}^{MB}_{thermal}|$ found from the visual, k-means and least squares methods, are, respectively, 9.3, 9.9 and 9.0 (Tables 13, 14 and 16). Each is within 10% or less than the others.

The undecomposed $\mathbf{Q}^{MB}$ found by the visual method is identical to the undecomposed $\mathbf{Q}$ from the FPI Standard Moments File found by adding together the FPI standard moments, $\mathbf{Q}_{bulk}$, $\mathbf{Q}_{enthalpy}$, and $\mathbf{Q}_{heatflux}$ (the sum of the small blue, red and the green vectors). The least squares and k-means methods *do not have* undecomposed $\mathbf{Q}^{MB}$ equal to the undecomposed $\mathbf{Q}$ from the FPI Moments File. Instead, each is equal to the undecomposed $\mathbf{Q}$ based on the standard moments of $f_{fit}(\mathbf{v})$ and $f_{box}(\mathbf{v})$.

`In Fig. (15), the three different pseudo-thermal energy density flux vectors, $\Delta\mathbf{Q}$, found from the three different methods are compared, The three $\Delta\mathbf{Q}$ vectors are all confined to a relatively narrow cone of angles. The smallest is the $\Delta\mathbf{Q}$ found by k-means, which therefore produces the smallest correction to $\mathbf{Q}_{thermal}$. However, $\Delta\mathbf{Q}$ is still a large correction to $\mathbf{Q}_{bulk}$. To show the corrections to this $\mathbf{Q}_{bulk}$ and to the $\mathbf{Q}_{bulk}$ vectors found by the other two methods we have added the $\mathbf{Q}_{bulk}$ vectors in cyan to the corresponding $\Delta\mathbf{Q}$ vectors, resulting in $\mathbf{Q}^{MB}_{bulk} = \mathbf{Q}_{bulk} + \Delta\mathbf{Q}$ for each method. In all cases $|\Delta\mathbf{Q}|$ is a large correction to $|\mathbf{Q}_{bulk}|$.

•

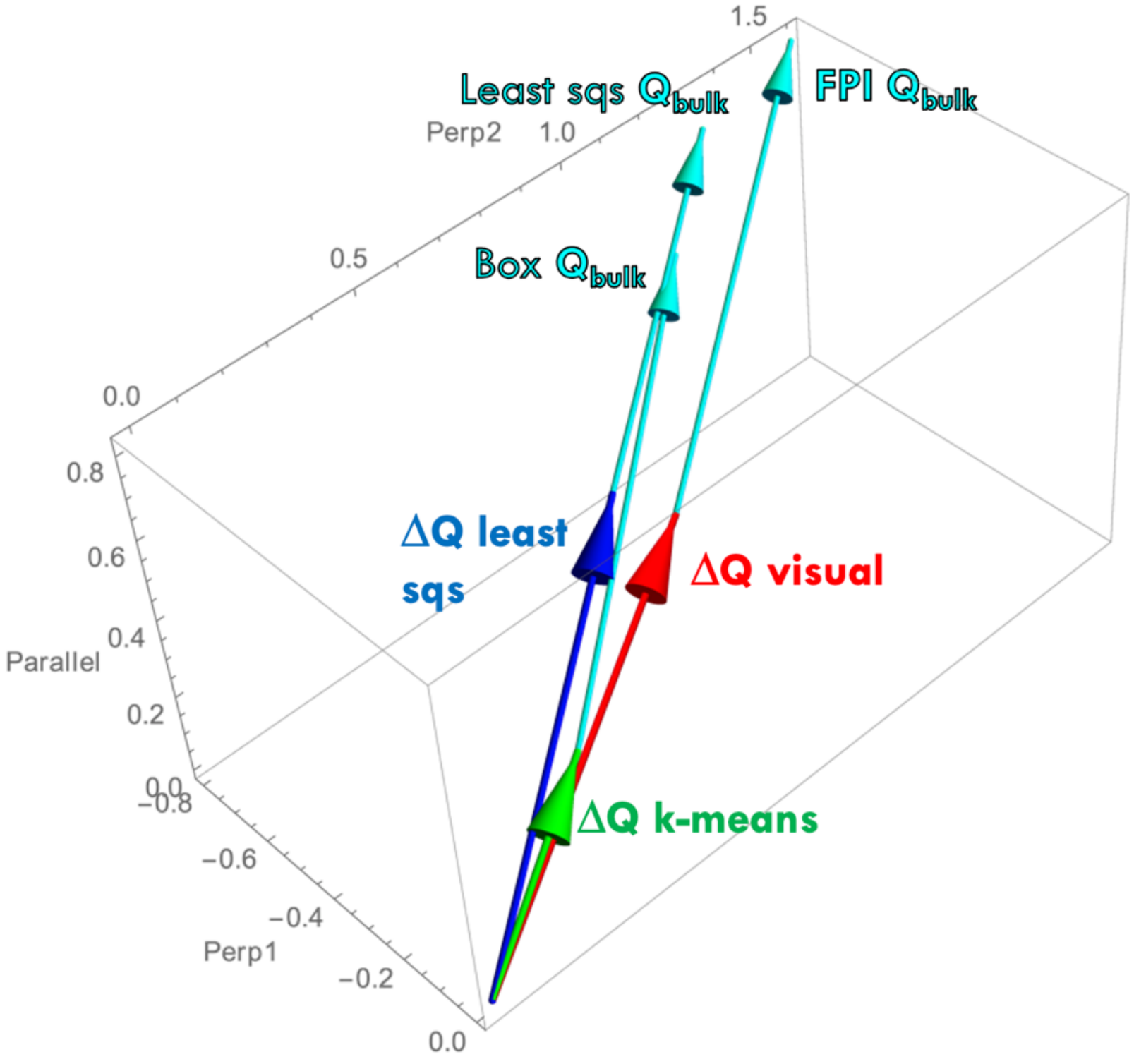

**Figure 15: Comparison of pseudothermal and bulk K. E. flux moment vectors**: $\Delta\mathbf{Q} = \mathbf{Q}_{therm} - \mathbf{Q}^{MB}_{therm} = \mathbf{Q}^{MB}_{bulk} - \mathbf{Q}_{buli}$ added to appropriate standard $\mathbf{Q}_{bulk}$ to yield $\mathbf{Q}^{MB}_{bulk}$ for each of the three methods. This enables a comparison of each standard bulk kinetic energy flux with the corresponding multibeam bulk kinetic energy flux. See Tables 10, 11, 12, 13, 16 and 14

## 7. Overall Summary and Significance

After reviewing concepts and methods developed in this paper and in the earlier paper, Goldman, et al, (2020), we summarize key results and discuss implications for emerging new measurements of multi-peaked and other compound velocity distributions.

*Standard* energy moments (Table 1) of compound ion velocity distributions, f(**v**), can contain *pseudothermal* parts, which may be misinterpreted as true temperature, pressure, etc. Pseudothermal parts are equivalent to deficits in the standard moments of both bulk kinetic energy and kinetic energy flux. Compound velocity distributions include f(**v**) with pronounced multiple peaks ("beams") and f(**v**) with extended two and three dimensionally shaped peaks ("ridges"). An alternate way of taking (multi-beam) energy moments of f(**v**), with *no pseudothermal* parts, is achieved by first approximating f(**v**) as a sum of N "beams," $f(\mathbf{v}) = f_1(\mathbf{v}) + ... + f_N(\mathbf{v})$. (Goldman, et al, 2020). Reasonable values of N are low single digits (Dupuis, et al, 2020)

### 7.1 METHODS

Three methods developed for approximating a compound f(**v**) as a sum of N beams (Goldman, et al, 2020) are applied to an f(**v**) measured in a PIC-simulation and to an MMS-measured f(**v**). They are

- A *visual method,* in which one estimates the N beam centroid (flow) velocities
- A *k-means method* in which f(**v**) is divided into N non-overlapping *clusters*
- A *least-squares method* in which each beam, $f_j(\mathbf{v})$, is found from a least-squares parametrized fit, $f_{fit}(\mathbf{v}) = [f_1(\mathbf{v}) + ... + f_N(\mathbf{v})]$ to f(**v**)

In the **visual method, multi-beam moments** of f(**v**) are constructed as follows:

- One visually estimates the N beam velocity centroids, $\mathbf{u}_j$, by referring to a 2D or 3D graphic display of f(**v**) (i.e., reduced f(**v**) planes, 3D fly-around, etc.). Optimum values of N are four or less, although five beams can be used if one has both a known centroid velocity *and* a known beam density (see *pedestals* in Sec. 3).
- The N beam-densities, $n_j$, are found from N ($\leq$ 4) velocities, $\mathbf{u}_j$, the standard flow velocity, **u**, and the standard density, n by matrix inversion. (Section 3).
- The densities and velocities are then used (via Table 2), together with *known standard* moments to find multi-beam moments.

***Strengths of the visual method*:**

- Intuitive
- Fast, once the centroids are specified.
- Only need velocity space image of the distribution.
- Multi-beam moments of f(v) can be found with little more than matrix inversion.
- Table 2 directly gives pseudothermal parts of a given set of standard thermal moments.
- Standard flow velocity, u, and density n of f(v) are pre-specified, so electron or ion beam fluxes are guaranteed to add up to n**u.**

***Limitations of the visual method*:**

- Must *estimate* beam centroid-velocities, $\mathbf{u}_j$ from 2D or 3D images of f(**v**) while avoiding choices leading to negative beam densities or negative multi-beam pressure eigenvalues.
- Number, N, of beams cannot be too large. Most straightforward approach requires $N \le 4$.
- Only beam densities and centroid velocities are determined explicitly. Beam shapes are never fully determined.
- Multi-beam *enthalpy flux*, $\mathbf{Q}^{MB}{}_{enthalpy}$, and *heat flux*, $\mathbf{Q}^{MB}{}_{heatflux}$ cannot be found separately; in the visual method, only their sum is found: the *thermal* energy flux, $\mathbf{Q}^{MB}{}_{thermal} = \mathbf{Q}^{MB}{}_{enthalpy} + \mathbf{Q}^{MB}{}_{heatflux}$.

**In the *k-means and least- squares* methods** multi-beam moments of f(**v**) are determined by taking a standard moment of each of the N beams in $f_1(\mathbf{v}) + ... + f_N(\mathbf{v})$ and adding. (Goldman, et al, 2020). N can be any integer. Beam shapes are fully determined.

***Strengths of k-means method***

- Method is systematic once N (= k) is chosen, and may be readily automated
- Method yields a lowest energy partition of f(**v**)
- With high energy cloud k-means works well with a *pedestal* (Sect. 4.4)

***Limitations of k-means method***

- The *beams* it yields are non-overlapping, non-interpenetrating *clusters* arising from the *partitioning* of f(**v**)

- The standard moment is based on the $f_{box}(\mathbf{v})$ used to generate particles rather than being the standard moment from the FPI Moments File.
- When there is a cloud, the straightforward k-means method tends to split it into pieces which are lumped in with the core beams. (This is remedied by splitting off the cloud as a *pedestal* and taking its standard moment before applying k-means to the remaining core (Sec. 4.4).)

***Strengths of least-squares method***

- Method is systematic once N is chosen, and method may be automated
- Beams can overlap and interpenetrate unlike in the k-means method
- Method can be modified to treat high energy velocity space *clouds* in f(**v**)

***Limitations of least squares method***

- Method requires choosing parametrized beam functions of velocity which may require modifications when high energy clouds are present in f(**v**).
- Standard moments are an output, not an input. They are moments of a fitted distribution, $f_{fit}(\mathbf{v})$, and may differ from the standard moments of the measured f(**v**), especially when high energy *clouds* are present in f(**v**).

**Additional consequences of high-energy *clouds* in the MMS f(v)**

- There are significant differences in the relative magnitudes of the energy density and energy density flux moments associated with the cloud in the MMS f(v) in comparison with those associated with the PIC f(**v**) (Table 17, below).
- High-energy *clouds* in the MMS f(**v**) contribute significantly to the heat flux moment.
- For each different method of finding multi-beam moments of MMS distribution must compare with standard moment of a slightly different f(v) (Table 9)
- The cloud in $f_{fit}(\mathbf{v})$ may be better represented by more than one beam in the least squares method of finding multibeam moments. (Table 15). This is especially true when the beams in the fit are taken to be symmetric functions of velocity and therefore lack heat flux moments.
- k-means works best with the cloud treated as a *pedestal* (Sect. 4.4) for MMS distribution

## 7.2 QUANTITIVE FINDINGS REGARDING MOMENTS

In this paper we have taken multibeam moments (Goldman, et al, 2020) of a compound ion velocity distribution, f(**v**), in a *PIC simulation* of magnetotail reconnection and of another f(**v**) taken from an *MMS-satellite-measured* f(**v**) during reconnection in the dayside magnetosphere. The multibeam *thermal energy density* moment, $U^{MB}_{therm}$ and the multibeam *thermal energy density flux vector* moment, $\mathbf{Q}^{MB}_{therm} = \mathbf{Q}^{MB}_{enthalpy} + \mathbf{Q}^{MB}_{heatflux}$, avoid the *pseudo-thermal parts* of the *standard* moments, $U_{term}$ and $\mathbf{Q}_{therm}$, but retain any *truly* incoherent thermal parts contained in the standard moments.

**Table 17: Comparison of multibeam energy and energy flux moments of PIC f(v) and of MMS f(v)**

| Issues | PIC simulation of tail reconnection | | MMS FPI measurement during dayside reconnection | |
|---|---|---|---|---|
| Is there a significant high energy cloud? | No. | | Yes | |
| How large are the ion pseudo-thermal energies, ΔU as a fraction of $U_{therm}$? | visual method | **60%** | visual method | **12%** |
| | k-means | **60%** | k-means | **8%** |
| | least squares | **47%** | least squares | **12%** |
| How large a fraction of the standard thermal flux is the heat flux, $\lvert\mathbf{Q}_{heatflx}\rvert / \lvert\mathbf{Q}_{thermal}\rvert$? | PIC $f_{box}(\mathbf{v})$ standard moment | **5%** | FPI standard moment file | **25%** |
| Magnitudes $\lvert\mathbf{Q}\rvert$, (determined by different velocity distributions) (Table 9) | visual method | **4.86** | visual method | **11.22** |
| | k-means | **4.95** | k-means | **11.32** |
| | least squares | **4.73** | least squares | **10.61** |

The pseudo-thermal energy density is defined as, $\Delta U = U_{therm} - U^{MB}_{therm} = U^{MB}_{bulk} - U_{bulk}$, and the pseudothermal energy density flux vector as, $\Delta\mathbf{Q} = \mathbf{Q}_{therm} - \mathbf{Q}^{MB}_{therm} = \mathbf{Q}^{MB}_{bulk} - \mathbf{Q}_{bulk}$. The pseudothermal part of the *standard* thermal energy density of f(**v**) which has been removed in the *multibeam* thermal energy density shows up as multibeam *bulk-kinetic* energy density), so that the sum of thermal and bulk kinetic energy density moments is the same whether the moments are standard or multibeam. In other words, $U^{MB}_{therm} = U_{therm} + \Delta U$, and $U^{MB}_{bulk} = U_{bulk} + \Delta U$. Likewise, $\mathbf{Q}^{MB}_{therm} = \mathbf{Q}_{therm} - \Delta\mathbf{Q}$ and $\mathbf{Q}^{MB}_{bulk} = \mathbf{Q}_{bulk} + \Delta\mathbf{Q}$.

One implication of our finding is that the traditional *standard* way of taking moments of a compound ion distribution, f(**v**), at a given location and time, can *exaggerate* the standard thermal energy density, $U_{therm}$, of f(**v**), and hence *exaggerate* the *ion temperature*, $T_i = U_{therm}/n$ at that location and time.

Multibeam energy moments therefore help sort out how much of a fluid element's energy at a given place and time is random (characterized by a true temperature) and how much of it is directed kinetic energy. This is significant because a set of *multibeam* energy moments taken at nearby places and times could help distinguish *particle heating* from *particle acceleration*,

A key issue is: by how much do *multibeam* thermal moments differ quantitatively from *standard* thermal moments, both for the PIC f(**v**) analyzed in Sec. 5 and for the MMS f(**v**) analyzed in Sec. 6? In other words, how large are pseudothermal effects in the two cases and can they be compared?

In Fig. 6 the standard and multibeam thermal energy densities for the PIC f(**v**) found by the various methods were compared (see units in Table 2). In most cases four beams were assumed in taking multibeam moments. In all cases the multibeam thermal energy densities are roughy half the standard thermal energy densities (also see Table 17). By contrast, the standard and multibeam thermal energy densities for the MMS f(**v**) in Fig. 13 are fairly close to each other.

From Table 17 we can get some idea of differences between the energy density moments of the PIC simulation f(**v**), which does *not* have a significant high-energy ion "*cloud*" and the MMS-measured f(**v**) which *does* have a significant high-energy ion cloud. For the PIC f(**v**), the ion pseudothermal energy density, $\Delta U_{therm}$, (based on four ion beams) was found to be around 50% - 60% of the standard energy density, $U_{therm}$, whereas for the MMS velocity distribution the ion pseudothermal energy density, $\Delta U_{therm}$, it was around 8% - 12% of the standard energy density, $U_{therm}$ . A probable explanation is that the cloud possesses a large amount of *true* thermal energy density that dominates both the standard and multibeam thermal energy density moments. In this scenario the (smaller) pseudothermal energy density in the MMS case would arise mainly from the lower-energy, colder beams.

Figures 7 and 8 show the standard and multibeam thermal and bulk-kinetic energy density *flux vectors* found by the three different methods, *as applied to the PIC f(**v**)*. The directions of the vectors are not significantly different from each other but the magnitude of the *standard* bulk kinetic energy density flux vector, $\mathbf{Q}_{bulk}$ (thick blue vector) is significantly smaller than the magnitudes of the corresponding (thin blue) *multibeam* vectors, $\mathbf{Q}^{MB}{}_{bulk}$, found by the three methods. In addition, the magnitude of the *standard* thermal energy density flux vector, $\mathbf{Q}_{therm}$, (in black in Fig. 7) is much larger than the corresponding *multibeam* vectors, $\mathbf{Q}^{MB}{}_{therm}$.

Figures 14 shows the standard and multibeam thermal and bulk-kinetic energy density *flux vectors* found by the three different methods as applied to the MMS f(**v**). The differences here are related to the high-energy ion velocity cloud in the MMS f(**v**), which was absent from the PIC f(**v**). The main difference is that the MMS standard heat flux, $\mathbf{Q}_{heatflux}$ is no longer negligible in comparison to the standard enthalpy flux, $\mathbf{Q}_{enthalpy}$ in the sum, $\mathbf{Q}_{therm} = \mathbf{Q}_{enthalpy} + \mathbf{Q}_{heatflux}$. This (frame-dependent) result is quantified in Table 17, which shows, for the PIC case, that, $|\mathbf{Q}_{heatflux}|$ is 5% of $|\mathbf{Q}_{thermal}|$, so that $|\mathbf{Q}_{thermal}|$ is dominated by $|\mathbf{Q}_{enthalpy}|$. However, for the MMS case, $|\mathbf{Q}_{heatflux}|$ is significant: It is equal to 25% of the magnitude of $|\mathbf{Q}_{thermal}|$. The heat flux can only be enhanced by a cloud if the cloud is highly asymmetric or if its velocity centroid is sufficiently skewed from the standard flow velocity. This asymmetry is prominently visible in Fig. 9.

In Figure 15 MMS pseudothermal flux vectors, $\Delta\mathbf{Q}$ found by the three methods (with four beams assumed or each method). Each of the $|\Delta\mathbf{Q}|$ is seen to be a significant fraction of the corresponding standard bulk kinetic energy density flux.

Finally, we turn to the pressure tensor moments, **P**, found by the three different methods for the PIC f(**v**) and for the MMS f(**v**). The standard and multibeam *pressure ellipsoids* are shown in velocity-space for the PIC f(**v**) and MMS f(**v**) cases in Figs. 4, 5 and 1. For the PIC f(**v**) the multi-beam ellipsoids differ in both size (they are smaller) and orientation from the standard ellipsoid, whereas for the MMS f(**v**) the multibeam ellipsoids are tightly embedded in each standard ellipsoid with the same orientation.

### 7.3 PRACTICAL SIGNIFICANCE OF MULTIBEAM MOMENT ANALYSIS

How might multibeam moments of compound velocity distributions impact studies of magnetic reconnection in the magnetosphere and elsewhere?

First, we point out that the potential impact is not limited to compound ion velocity distributions but extends to electron distributions as well. For example, during reconnection, electron *crescents* have been measured together with electron core distributions which may consist of both cold and hot components. We have presented multibeam electron moment studies of compound electron f(**v**) containing crescents at Workshops and Meetings (e.g., AGU, 2019).

Finding reliable moments of compound electron and ion distributions is germane to the identification of particle diffusion regions during reconnection as well as to particle heating and acceleration. Off-diagonal pressure tensor elements, for example may break the frozen-in condition of electrons or ions, since the pressure force enters into the particle momentum equation together with electric and magnetic forces (Aunai, et al 2011; Dai, et al, 2015).

Our results show that pseudothermal effects are less pronounced for the MMS ion f(**v**) than for the PIC f(**v**) due to a high energy ion velocity-space cloud present in the former but missing from the latter. For example, the MMS f(**v**) has a higher magnitude ion heat flux, $|\mathbf{Q}_{heatflux}|$ relative to that of the ion enthalpy flux, $|\mathbf{Q}_{enthalpy}|$, when compared to the size of $|\mathbf{Q}_{heatflux}|$ in the PIC f(**v**). This might appear to suggest that the PIC simulation should have included higher-energy ion velocities to properly evaluate higher order moments such as the ion heat flux.

A possible alternate explanation is the following: The PIC simulation is based on *magnetotail* reconnection measurements while the MMS f(**v**) is from a dayside *magnetopause* reconnection event. Measurements by Eastwood, et al (2013) of energy flux moments during antiparallel symmetric reconnection in the magnetotail suggest that the energy flux is dominated by ion enthalpy, with contributions from the electron enthalpy and heat flux and ion kinetic energy flux. The electron kinetic energy and ion heat flux appear to be negligible.

Even if they are present during magnetotail reconnection high-energy ion clouds can be less relevant to heat flux moments if the clouds are symmetric and centered at the standard flow velocity. The PIC simulation studied here may not be a bad approximation. At any rate, the magnetotail should be an interesting future arena for carrying out multibeam analyses of compound ion distributions.

The three different methods (visual, k-means and least-squares-fit) used in this paper to identify the ion beams whose sum is an approximation to a measured compound ion f(**v**) have enabled multibeam moments of a measured f(**v**) to be taken practically and perhaps eventually automated (k-means is the best candidate). By comparing the multibeam thermal energy density moments, $U^{MB}_{therm}$ to the standard moment, $U_{therm}$, over a neighborhood of nearby space-time locations, one may ultimately be able to distinguish between whether a species has undergone *heating* or has undergone *particle acceleration*. Hopefully, the multibeam methods presented here will also be instructive for future applications to compound particle velocity distributions measured in the magnetosphere and to emerging measurements of compound distributions in the solar wind (e.g. from the Solar Orbiter and the Parker Solar Probe missions).

**Acknowledgments**

We wish to thank Dr. Steven Schwartz for helpful conversations leading to this paper. JPE is supported by UKRI (STFC) grant ST/ S000364/1. M. Goldman, D. Newman and G. Lapenta are supported by NASA Grants NNX08AO84G, 80NSSC17K0607 and 80NSSC19K0841. We thank NASA for the computational resources used to generate the graphics. The MMS data is accessible through the public link provided by the MMS science working group teams: http://lasp.colorado.edu/mms/sdc/public/. We are grateful for the dedicated efforts of the entire MMS mission team, including development, science operations, and the Science Data Center at the Univ. of Colorado.